\documentclass[aps, twocolumn, longbibliography,amsmath,amssymb, superscriptaddress, prx]{revtex4-2}

\usepackage{MnSymbol}
\usepackage{graphicx}
\usepackage{subfigure}
\usepackage{epstopdf}
\usepackage{hyperref}
\usepackage{bm}
\usepackage{color}
\usepackage{braket}
\usepackage{lipsum}  
\usepackage{xcolor}
\usepackage{comment}
\usepackage{listings}
\usepackage{cancel}
\usepackage{mathrsfs}
\usepackage{amsmath, amssymb}

\begin{document}

\title{Fractional Spin Quantum Hall Effect in Weakly Coupled Spin Chain Arrays}
\author{Even Thingstad}
\affiliation{Department of Physics, University of Basel, Klingelbergstrasse 82, CH-4056 Basel, Switzerland}
\author{Pierre Fromholz}
\affiliation{Department of Physics, University of Basel, Klingelbergstrasse 82, CH-4056 Basel, Switzerland}
\author{Flavio Ronetti}
\affiliation{Department of Physics, University of Basel, Klingelbergstrasse 82, CH-4056 Basel, Switzerland}
\affiliation{Aix Marseille Univ, Universit\'e de Toulon, CNRS, CPT, Marseille, France}
\author{Daniel Loss}
\affiliation{Department of Physics, University of Basel, Klingelbergstrasse 82, CH-4056 Basel, Switzerland}
\author{Jelena Klinovaja}
\affiliation{Department of Physics, University of Basel, Klingelbergstrasse 82, CH-4056 Basel, Switzerland}

\begin{abstract}
Topological magnetic insulators  host chiral gapless edge modes. In the presence of strong interaction effects, the spin of these modes may  fractionalize. Studying a 2D array of coupled insulating spin-$1/2$ chains, we show how spatially modulated magnetic fields and Dzyaloshinskii-Moriya interactions can be exploited to realize chiral spin liquids or integer and fractional spin quantum Hall effect phases. These are characterized by a gapped bulk spectrum and gapless chiral edge modes with fractional spin. The spin fractionalization is manifested in the quantized spin conductance, which can be used to probe the fractional spin quantum Hall effect. We analyze the system via bosonization and perturbative renormalization group techniques that allow us to identify the most relevant terms induced by the spin-spin interactions that open gaps and render the system topological under well-specified resonance conditions. We show explicitly that the emerging phase is a genuine chiral spin liquid. We suggest that the phases can be realized experimentally in synthetic spin chains and ultracold atom systems.
\end{abstract}

\maketitle

\section{Introduction}

Entanglement enables remarkable trade-offs in the arrangement of low-energy eigenstates within frustrated many-body quantum systems. While first introduced by Anderson~\cite{Anderson1973} through the resonating valence bond state in antiferromagnetic systems, quantum spin liquids (QSL)~\cite{Balents2010_Spin, Mourigal2013_Fractional,  Savary2016_Quantum, Zhou2017_QuantumSpinLiquid, Hermanns2018_PhysicsKitaev, Knolle2019_FieldGuide,Broholm2020}  now exist in many flavours. 
They are characterized by the absence of long-range magnetic order down to zero temperature due to quantum fluctuations. 
They require strong frustration~\cite{Rokhsar1988,PhysRevLett.66.1773,PhysRevLett.86.1881,PhysRevB.62.7850,PhysRevB.73.155115,Kitaev2006_Anyons,Lee2008,Lee2006_Doping,Balents2010_Spin} resulting in highly degenerate many-body entangled states capable of hosting chargeless non-local excitations~\footnote{I.e. excitations created by an infinite product of local operators}. These non-local excitations are adiabatically connected to partonic
or fractional deconfined quasi-particles~\cite{KITAEV20032,Nayak2008}, much like spinonic or solitonic domain walls in 1D Heisenberg chains~\cite{Kohno2007,Mourigal2013_Fractional}. 
The topological protection of these excitations provides two advantages. First, their robustness enables reliable transport properties. 
Second, they display anyonic statistics~\cite{PhysRevB.25.2185, Haldane1983_Fractional, Wilczek1990,PhysRevLett.49.957,PhysRevLett.51.2250,Arovas1985,Frhlich1988,Frhlich1989,PhysRevLett.95.246802,PhysRevB.72.075342}, which could be useful to implement and to manipulate qubits~\cite{Nayak2008}.

While QSLs are challenging to understand, one may draw on parallels to other strongly correlated quantum liquids, for instance describing superconductors~\cite{Bednorz1986_Possible,anderson1988frontiers,Laughlin1988} and the fractional quantum Hall effect (FQHE)~\cite{Tsui1982_TwoDimensional, Laughlin1990_ElementaryTheoryIncompressibleQuantumFluid, Laughlin1983_Anomalous, Stone1992, Halperin1984_Statistics, PhysRevLett.53.722, Prange1990}. 
Kalmeyer and Laughlin exploited the parallel to the latter~\cite{PhysRevLett.59.2095,Kalmeyer1989} and suggested that the resonating valence bond state of Anderson could be described by a variant of
Laughlin's wave function~\cite{Laughlin1983_Anomalous}.
The parent phase is now known as a chiral spin liquid (CSL) and is characterized by a gapped bulk, gapless chiral edge states, and fractional modes with fractional statistics. 
By definition, CSLs have topological order~\cite{PhysRevB.71.045110, PhysRevB.81.060403,Isakov2011,PhysRevB.40.7387, PhysRevB.41.9377, Wen2007b, Pereira2020} and are classified by the projective special linear 
group~\cite{PhysRevB.87.125127,PhysRevB.93.094437}. 
Crucially, both the FQHE and the CSL involve breaking of the time-reversal ($T$) and parity ($P$, i.e. mirror symmetry in two-dimensional spin systems) symmetries but not of the product of the two~\cite{PhysRevB.61.10267}. Indeed, under a $P$ or $T$ transformation, particles with statistical angle $\theta$ change into particles with statistical angle $-\theta$~\cite{PhysRevLett.61.2066}. Breaking the $P$ and $T$ symmetries 
allows the existence of chiral edge modes without their $P$- or $T$-symmetric partner of opposite chirality.
This suppresses backscattering, which is the prime obstacle for transport at low energy.
In the FQHE, a strong magnetic field violates the $P$ and $T$ symmetries.
In the CSL context, the symmetry breaking must be of a different origin~\footnote{A constant magnetic field alone generates magnetic ordering, and additional isotropic nearest-neighbour spin exchange seems insufficient to obtain a CSL.}. 

\begin{figure}
    \centering
    \hspace*{-0.2cm}  
    \includegraphics[width=0.92\columnwidth]{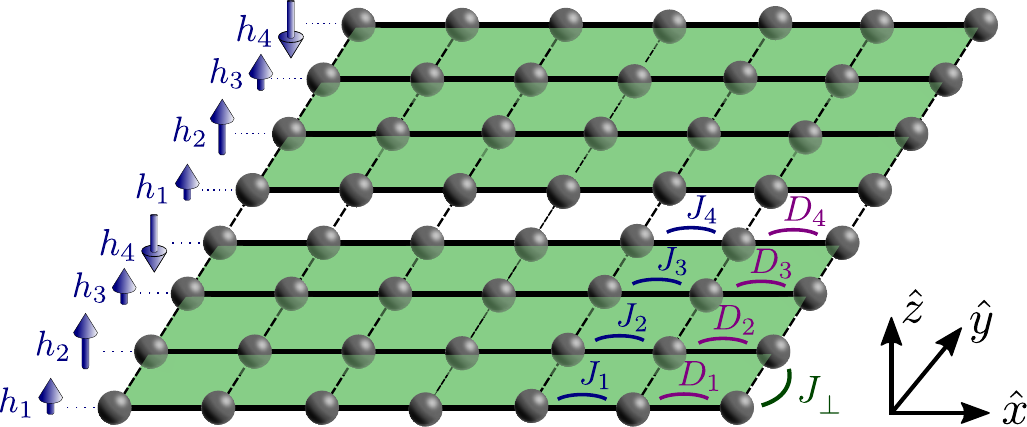}
    \caption{Sketch of model. Two-dimensional spin lattice consisting of localized spin-$1/2$'s (spheres) in a repeated pattern of coupled chains (solid lines) with chain-dependent intrachain exchange ($J_l$) and Dzyaloshinskii-Moriya interactions ($D_l$), as well as interchain exchange interactions ($J_\perp$). By tuning the chain magnetizations along $z$ independently through external magnetic fields \(h_l\) (arrows) applied along $z$-direction, one may bring the system into the fractional spin quantum Hall effect regime with magnetic excitations carrying fractional spin and with chiral edge states. These phase are also identified as chiral spin liquids.}
\label{fig_model}
\end{figure}

Several strategies exist to write down a model system hosting a CSL. One approach is to start from a spin (or hard-core bosonic) version of Laughlin's Ansatz for half-integer fractional modes~\cite{WANG2008} and then to construct a parent Hamiltonian with the chiral spin state as its ground state~\cite{PhysRevLett.59.2095, Kalmeyer1989,PhysRevLett.99.097202, PhysRevLett.108.257206, PhysRevB.89.165125, PhysRevB.80.104406, PhysRevLett.70.2641, PhysRevB.41.664}. Hamiltonians obtained this way are typically very challenging to implement experimentally since they contain contrived 
long-range interactions. Alternatively, one may write down a Hamiltonian with explicit breaking of $P$ and $T$ symmetry. In the literature, this is 
achieved by considering the frustrating triple-spin interaction $\chi_{ijk} = \bm{S}_i\cdot (\bm{S}_j \times \bm{S}_k)$~\cite{PhysRevB.39.11413, PhysRevLett.63.2524, PhysRevB.95.144413}.
Here \(i,j,k\) denote lattice sites, typically on triangular or kagomé lattices~\cite{PhysRevLett.99.247203,Bauer2013,Bauer2014,PhysRevB.95.035141,PhysRevB.94.075131, PhysRevB.97.195158, PhysRevB.98.184409}, although in general, any lattice may be used~\cite{PhysRevB.39.11424,PhysRevLett.70.2641, PhysRevB.85.155145}.
The chiral interaction $\chi$ may occur in the effective spin model describing a Mott insulator through 
a strong magnetic field~\cite{Motrunich2006_Orbital, PhysRevB.51.1922} or circularly polarized light~\cite{Claassen2017_Dynamicical}. 
Finally, the $P$ and $T$ symmetries can break spontaneously. 
Indeed, CSLs have been identified in antiferromagnetic Heisenberg models with longer-ranged frustrating interactions~\cite{PhysRevB.39.11413, PhysRevLett.63.2524, PhysRevLett.112.137202, Gong2014, PhysRevB.91.075112, Yan2011,  PhysRevB.92.125122, PhysRevB.92.220404,PhysRevLett.114.037201,PhysRevLett.108.207204,PhysRevB.97.241110}. Moreover, these frustrating interactions can also contribute to stabilizing CSL phases in models with explicitly broken symmetry~\cite{PhysRevB.91.041124, PhysRevB.92.125122,PhysRevB.94.075131}.

Many strongly interacting systems can conveniently be understood through a ``coupled wire approach''~\cite{Kane2002_Fractional,Teo2014_From, Meng2020_Coupled}. While originally designed for the FQHE in two-dimensional electron gases~\cite{PhysRevLett.58.270, PhysRevB.43.11353,PhysRevB.63.054430, Klinovaja2014_Integer, Neupert2014_WireDeconstructionism, PhysRevX.4.011036,PhysRevB.87.045128, PhysRevB.89.104523, Meng2014, Sagi2014_NonAbelianTiFromArrayOfWires, PhysRevB.90.235425, PhysRevB.91.245144, Laubscher2021_FractionalBoundaryCharges, Klinovaja2014_Integer, Klinovaja2013_TopologicalEdgeStatesFromUmklapp, Klinovaja2015_IntegerFractionalQuantumAnomalousStripOfStripes, Laubscher2021_FractionalBoundaryCharges}, it is useful also for fractional topological insulators~\cite{Sagi2014_NonAbelianTiFromArrayOfWires,PhysRevB.90.115426, PhysRevB.90.235425, Santos2015_FractionalTiSlidingLlToCS, Neupert2014_WireDeconstructionism, Sagi2014_NonAbelianTiFromArrayOfWires, Rachel2018_InteractingTiReview, PhysRevB.86.125119}, liquids of interacting anyons~\cite{Ludwig2011,PhysRevLett.103.070401}, ladder models~\cite{Oreg2014_Fractional,PhysRevB.89.045111,PhysRevB.91.115427}, high-$T_c$ superconductors~\cite{PhysRevB.62.3422,PhysRevLett.85.2160}, bilayer or Moir\'e materials~\cite{PhysRevB.88.121408,PhysRevLett.126.186601,PhysRevB.104.045146,PhysRevLett.113.236804}, fractional chiral metals~\cite{Meng2016_TheoryOf31FracChiralMetal}, higher-order topological phases~\cite{Laubscher2023_FractionalSecondOrder, Meng2015_FractionalTopPhases}, and also CSLs~\cite{Gorohovsky2015, Meng2015_Coupled, Huang2016_NonAbelian}.
The starting point of the coupled wire approach is unpaired gapless Tomonaga-Luttinger liquids. 
Their excitations are in general non-integer~\cite{Pham2000} but not robust, since a local perturbation will change their charge, spin, or statistics.
By coupling one wire to the next in a way that breaks the $P$ and $T$ symmetries, one may introduce a gap-opening backscattering between left- and right-moving modes on subsequent wires. This results in a topologically protected bulk gap and chiral edge modes. The elementary bulk excitations are kinks associated with each bulk backscattering. Depending on the details of the interwire interaction, these may be anyonic. This scenario can be described by effective field theories involving a collection of sine-Gordon models or SU(2)$_k$ Weiss-Zumino-Novikov-Witten models, which both describe the elusive fractional modes.  
The initial one-dimensional formalism then helps identify the ground state degeneracy~\cite{PhysRevB.41.9377} as well as the bulk and edge modes~\cite{PhysRevB.43.11025} associated with the observable and characteristic features of CSLs. These include quantized fractional heat and spin conductivity, potential quantized fractional heat and spin \textit{Hall} conductivity, and deconfined~\cite{Kalmeyer1989,PhysRevLett.66.276,PhysRevLett.70.2641} fractional spin excitations with Abelian or non-Abelian braiding statistics~\cite{PhysRevLett.102.207203}.

The identification of CSL phases is challenging both in simulations~\cite{PhysRevB.84.075128,PhysRevLett.107.077203,PhysRevLett.107.087205,PhysRevB.84.140404} and experiments. In simulations, finite-size effects can prevent consensus on the nature of the ground state~\cite{Yan2011, PhysRevLett.109.067201, PhysRevB.87.161118}. In experiments, one may consider realizations with hard-core bosons~\cite{PhysRevLett.114.037203}, bosons~\cite{PhysRevB.79.165301} or fermions~\cite{PhysRevB.91.134414,PhysRevLett.116.137202}.
The most popular candidates are kapellasite and related materials~\cite{PhysRevLett.101.106403, PhysRevLett.109.037208,PhysRevB.87.155107,PhysRevB.90.205103,PhysRevB.91.220408} or $\alpha$-RuCl$_3$, which have thermal conductance compatible with CSL behaviour~\cite{Yamashita2008, Hirschberger2015_Thermal, Hirschberger2015_Science, Katsura2010, Kasahara2018_Majorana, Hentrich2019_Large}. Magnetic torque measurements~\cite{PhysRevB.98.205110,PhysRevB.99.081101} could potentially provide additional indications~\cite{PhysRevB.99.174429}, but undisputed observation of the CSL phase remains elusive. This can be attributed to at least three factors, common to all QSLs.
First, QSLs may only appear when different spin interactions are competing~\cite{PhysRevLett.93.127202}. Hence, the QSL phase competes with broken symmetry phases such as spin density wave or valence bond crystal phases~\cite{Iwase1996,PhysRevLett.73.3463,Meng2010,PhysRevB.86.024424} and may be restricted to a tiny parameter space region, or even a fine-tuned critical point~\cite{PhysRevB.67.024422,Senthil2004_Quantum_PRB}. Second, magnetic susceptibility measurements can only demonstrate the absence of magnetic order~\cite{PhysRevB.99.174429, Hiroi2001,Ofer2006,PhysRevLett.98.107204,PhysRevLett.100.087202,Yamashita2008,Okamoto2009,PhysRevB.83.214406}. Finally, the effect of interactions beyond minimal models can be very destructive to the CSLs, such as, for example, the interaction with phonons in solid state systems~\cite{PhysRevX.8.031032}.

These challenges emphasize the need for different model systems hosting CSLs. Within the coupled wire approach, the interchain interactions usually involve the complicated $\chi$ interaction term explicitly~\cite{Gorohovsky2015, PhysRevLett.123.137202,PhysRevLett.115.267209, Huang2016_NonAbelian, PhysRevB.95.140406, PhysRevB.99.174429}
or effectively through a fine-tuned Rashba spin-orbit interaction~\cite{Meng2015_Coupled}. We propose to modulate simpler types of couplings to achieve fractional modes as well as $P$ and $T$ symmetry breaking~\cite{2018PhDT........63J,10.21468/SciPostPhysCore.2.2.004}. In two-leg spin ladders~\cite{Timonen1985_Continuum, Schulz1985_Phase, Nersesyan1998_Incommensurate, Nayak2020_MagneticField}, chain-dependent magnetic fields may produce fractional spin excitations~\cite{Ronetti2022_Fractional}, as also verified by density matrix renormalization group methods.
Chain-dependent Dzyaloshinskii-Moriya interactions (DMI) break the $P$ and $T$ symmetries. According to mean field arguments, this allows the emergence of an integer spin quantum Hall effect~\cite{Hill2017_Spin}. 

In this paper, we consider the two-dimensional  lattice spin model illustrated in Fig.~\ref{fig_model}. Modulating both the magnetic field and the DMI spatially, we break the $P$ and $T$ symmetries to obtain fractional modes.
Using a coupled-wire approach and bosonization allows us to capture spin fractionalization of magnetic excitations, as opposed to mean-field techniques. We show that our model hosts several different fractional spin quantum Hall effect (FSQHE) or CSL phases within the first and second levels of the hierarchy classification of FQHE phases. We calculate the spin conductance within these phases and show how it relates to the fractional spin and the integers specifying the FSQHE phases in the strongly interacting regime. 

This paper is organized as follows: In Sec.~\ref{sec_nutshell}, we provide a brief pedagogical introduction to CSLs. In Sec.~\ref{sec_model}, we present the model
and discuss how to bosonize it and generate interaction terms perturbatively. In Sec.~\ref{sec_fsqhe}, we discuss how to engineer the system so that it hosts a first-level FSQHE phase and discuss the weak coupling phase diagram, the fractional spin, and the spin conductance of this phase, as well as a local chiral order parameter. In Sec.~\ref{sec_higherLevel}, we discuss how to construct second-level FSQHE phases and their properties. In Sec.~\ref{sec_breakingMagnetizationConservation}, we discuss the effect of perturbations breaking the conservation of magnetization. In Sec.~\ref{sec_discussion}, we discuss possible experimental realizations and in Sec.~\ref{sec_summary} we summarize. 

\section{Chiral spin liquids in brief}\label{sec_nutshell}

We briefly overview here the essential features of a CSL. The phase of a spin system is a gapped CSL if it meets the following two criteria~\cite{PhysRevB.39.11413}: First, the ground state of the system 
is a chiral spin state, i.e. a state without magnetic order~\footnote{For which we use the usual definition of a zero spin-spin covariance when the two spins are far apart.} but with a non-zero local CSL order parameter \(\chi\).
Second, the phase is a liquid, meaning that the local order parameter can be extended to a non-local one, as expected from the topological order. 
In the following, we provide two different well-known perspectives on CSLs and these criteria, based on microscopic and field-theoretic descriptions of a CSL.  This characterization will then be used in the main part of this work to identify the emerging FSQHE phases of our microscopic model as CSLs.

\subsection{Explicit chiral spin state}\label{sec_chiral_state}

We now present a way to build an explicit chiral spin state~\cite{PhysRevB.39.11413}. 
We consider the spin-1/2 operators \(S_i^\alpha\) labelled by the index $i$
along the axis \(\alpha \in \{x, y, z\}\). Two sufficient conditions for a state to be a chiral spin state are zero net magnetization, \(\sum_i \langle  S_i^\alpha \rangle = 0\), and a non-zero average value \(\langle \chi_{ijk}\rangle\) for the chirality operator \( \chi_{ijk} = \bm{S}_i \cdot (\bm{S}_j \times \bm{S}_k )\). Here, the brackets denote the average value with respect to the chiral spin state. The chirality operator is considered local as long as the spins labeled by \(i\), \(j\), and \(k\) are close to each other~\footnote{Strictly speaking, such a notion of locality requires the definition of a lattice that we do not wish to specify at this stage for more generality.}.

Both conditions for the chiral spin state can be satisfied already with four spins. By the properties of the SU(2) Lie algebra, %there exist exactly two spin singlets out of the $2^4$ possible four-spin states. By definition, only these two states satisfy (a). 
two out of the $2^4$ four-spin states are spin singlets, and therefore have zero net magnetization. 
These are
\begin{equation}
\begin{split}
    \lvert S=0 \rangle^\alpha = \,\, & \lvert \uparrow \uparrow \downarrow \downarrow \rangle + \lvert  \downarrow \downarrow \uparrow \uparrow \rangle +\omega_\alpha \left(\lvert \uparrow \downarrow\uparrow  \downarrow \rangle + \lvert \downarrow\uparrow \downarrow\uparrow   \rangle \right) \\
    & \qquad \qquad \quad \,\,\,\,\,\, + \omega_\alpha^2  \left(\lvert \uparrow \downarrow \downarrow \uparrow  \rangle + \lvert  \downarrow \uparrow \uparrow \downarrow   \rangle \right),
\end{split}
\end{equation}
where $\alpha = \pm 1$ and $\omega_\alpha =e^{i 2\pi \alpha /3}$. 
Both states are also eigenvectors of the operator $\chi_{i j k}$ with eigenvalues $\pm  2 \sqrt{3}$, with sign depending on $\alpha$ and which three spins \((i,j,k)\) we consider. 
The states \(| S=0 \rangle^\alpha\) are therefore chiral spin states. 

The chiral four-spin states can be exploited to construct a macroscopic spin state that also satisfies both criteria. 
For this, we consider $4N$ spins on a lattice and partition them into disjoint groups of four nearby spins. The tensorial product of any of the two states $\lvert S=0 \rangle^\alpha$
for each group is a separable
chiral spin state.
Using projectors, one may construct a local Hamiltonian with a particular tensorial product as its ground state~\cite{PhysRevB.39.11413}. Since the state is locally separable, however, it does not describe a liquid.

A chiral spin state describes a CSL only if it also has a finite non-local chiral order parameter. Such an order parameter can be constructed by considering a Wilson loop within a spinon description of the CSL~\cite{PhysRevB.39.11413}. Alternatively, the existence of {\it chiral edge states} ensures the breaking of chirality at large scales. Since a chiral spin liquid is a gapped topological phase, the bulk-edge correspondence also ensures the breaking of chirality in the bulk.
As is often the case in topological phases, obtaining an explicit expression for the microscopic ground state is very difficult. Instead, it is better understood from a low-energy field-theoretic point of view.

\subsection{Field theory for a chiral spin liquid}

The FQHE and its edge states can be described through a variety of effective field theories,
for instance, through coupled chiral Luttinger liquids~\cite{PhysRevB.41.12838,RevModPhys.75.1449}, SU(2)$_k$ Weiss-Zumino-Novikov-Witten models~\cite{PhysRevB.43.11025,BuenoXavier2023,PhysRevB.43.11025,PhysRevB.43.1257,PhysRevB.43.10622}, or more generally Chern-Simons theories~\cite{Witten1989}. 
By extension, there are many ways to describe CSLs.

Here, we present a toy model for a CSL.
The starting point is a system of $N$ independent spin-1/2 chains with isotropic intrachain spin coupling in the $xy$-plane and no interchain interactions. Taking the continuum limit and using Abelian bosonization, these chains can be described by a Luttinger liquid at low energy. The corresponding Hamiltonian is 
\begin{equation}\label{eq:luttinger}
    \mathscr{H}_0 = \sum_{l=1}^{N} \int \frac{dx}{2\pi} u \left[ (\partial_x \theta_l)^2 + (\partial_x \phi_l)^2 \right],
\end{equation}
where $u$ is the velocity scale, and \(\phi\) and \(\theta\) are the Abelian bosonic fields in the low-energy description of the system. To make the system a chiral spin liquid, we need to introduce interactions. Importantly, the interaction terms need to %be chosen so that they 
break the chiral symmetry %also 
in the effective low-energy description of the system obtained through Abelian bosonization. In the following, we first write down such a low-energy effective interaction, before discussing its relation to the chiral order parameter \(\chi_{ijk}\). Finally, we sketch an attempt to construct a microscopic interaction that would induce a CSL phase
to illustrate why realistically engineering such an interaction  
is challenging.

We first consider the effective low-energy interaction term
\begin{align}\label{eq_toyInteraction}
\mathscr{H}_\perp^\mathrm{eff} = g_\chi \sum_{l=1}^{N-1} \int \frac{dx}{2\pi} \, \cos [\eta_{l}^{R}(x)- \eta_{l+1}^{L}(x)],
\end{align}
where \(\eta_l^L=\theta_l + 2\phi_l\) and \(\eta_l^R = \theta_l - 2\phi_l\) are chiral fields. The Hamiltonian \(\mathscr{H} = \mathscr{H}_0 + \mathscr{H}_\perp^\mathrm{eff}\)
describes a sine-Gordon model and can be understood by renormalization group methods. When the interaction coupling strength $g_\chi \rightarrow \infty $
under the renormalization group flow, the interaction
dominates the physics.  In the ground state, the commuting fields \(\eta_{l}^{R}- \eta_{l+1}^{L}\) are %therefore 
pinned, and 
the bulk modes are gapped by the interaction. In contrast, $\eta_{1}^{L}$ and $\eta_{N}^{R}$ are completely decoupled from the bulk 
and represent topologically-protected gapless chiral edge modes. 

To demonstrate that the phase described above has local chiral order, we consider the bosonized form of \(\chi_{ijk}\), where \((i,j,k)\) denotes a set of nearby spins on the microscopic lattice. For concreteness, we consider coupled spin chains with sites forming a square lattice and let \(\bm{S}_{j,l}\) be the spin at position $x_j=j a$ of chain $l$, where $a$ is the lattice constant. We then pick out triplets of spins from a given smallest plaquette to construct a simple local chiral order parameter that is symmetrized with respect to the two chains it involves.
While there are numerous ways to do the symmetrization, we consider
\begin{align}\label{eq_toyModel_chiralOrderParam}
\chi_{l}(x_j) = (\bm{S}_{j,l} - \bm{S}_{j+1,l+1} ) \cdot \bm{S}_{j+1,l} \times \bm{S}_{j,l+1}.
\end{align}

We bosonize the spin operators in Appendix~\ref{app_bosoChi} and show that the bosonized chiral field \(\chi_l(x)\) consists of many terms, including terms of the form \(\cos (\eta_l^R - \eta_{l+1}^L)\).
When the interaction in Eq.~\eqref{eq_toyInteraction} dominates the physics, the local chiral order parameter therefore develops a finite expectation value. 

The most natural ansatz for a model spin interaction giving rise to a chiral spin liquid is \(\mathscr{H}_\perp^\chi = J_\chi \sum_{jl} \chi_l(x_j)\). Such a complicated three-spin interaction has been shown to lead to a chiral spin liquid for variants of a triangular lattice~\cite{Gorohovsky2015,PhysRevB.95.140406}. In contrast, we consider a microscopic spin model with only two-spin interactions and on a square lattice. We demonstrate the existence of several different CSL phases (in form of FSQHE phases) for this case, even though competing magnetic order is not suppressed by geometric frustration.

\section{From Spin Lattice to Bosonic Low-Energy Model}\label{sec_model}
We turn now to the main focus of this work and introduce a microscopic spin model that will support CSL phases, and, in particular, exhibit a 
FSQHE.
For this we first introduce a lattice  spin model and then analyze it via bosonization of the intrachain Hamiltonian, of the Jordan-Wigner string, and of the interchain Hamiltonian.
Finally, we discuss how the $P$- and $T$-breaking interaction responsible for the CSL phase can emerge within a perturbative approach.

\subsection{Lattice spin Hamiltonian}
We consider a setup consisting of \(N\) coupled spin chains described by  the Hamiltonian \(H = \sum_l {H}_l^0+H_\perp\), as illustrated in Fig.~\ref{fig_model}. 
The intrachain Hamiltonian \({H}_l^0\) describing chain \(l\) includes both nearest-neighbour spin exchange and Dzyaloshinskii-Moriya interactions and is given by
\begin{equation}
{H}_l^0 =  \sum_{j,\alpha} J^\alpha_{l} {S}^\alpha_{j,l} {S}^\alpha_{j+1, l} - D_l  \sum_{j}  \hat{z} \cdot (\bm{S}_{j,l} \times \bm{S}_{j+1, l})  - h_l \sum_j S_{j,l}^z,
\end{equation}
 where \(S_{j,l}^\alpha\) is the component \(\alpha\) of the spin-\(1/2\) operator associated with lattice site \(j\) of the \(l\)-th chain. Here, \(J_l^\alpha\) and \(D_l\) are the exchange and Dzyaloshinskii-Moriya coupling strengths, respectively.  
The magnetic fields \(h_l\), applied along the $z$-direction, tune the magnetization along the $z$-axis in the chains. The spins in different chains interact through an exchange interaction between spins in nearest-neighbour chains, and the interchain Hamiltonian is given by
\begin{align}\label{eq:hperp}
H_\perp = \sum_{j,l, \alpha} J_\perp^\alpha S_{j,l}^\alpha S_{j,l+1}^{\alpha},
\end{align}
with interchain exchange coupling strengths \(J_\perp^\alpha\). To respect U(1) symmetry and have conservation of the total magnetization along the external magnetic field, we assume \(J_l^{x} = J_l^y \equiv J_l \) and  \(J_\perp^x = J_\perp^y \equiv J_\perp^{xy}\). Furthermore, we consider \(J_l^{z}\geq 0\). In general, we allow \(J_\perp^z\) to have any sign but may later need \(J_\perp^z >0\) to make the FSQHE interaction relevant in the weak coupling regime.

The spin system can be mapped to a system of spinless particles through a Jordan-Wigner transformation~\cite{Jordan1928_UberPaulische}. The creation and annihilation operators \(c_{j,l}^\dagger\) and \(c_{j,l}\) describing these particles are related to the spins through 
\begin{subequations}
\label{eq_jordanWigner}
\begin{align}
S_{j,l}^x &= \,\,\,\, \frac{1}{2} (c_{j,l}^\dagger + c_{j,l} ) W_{j,l}   , \\
S_{j,l}^y &= -\frac{i}{2} (c_{j,l}^\dagger  - c_{j,l}) W_{j,l} , \\
S_{j,l}^z &= c_{j,l}^\dagger c_{j,l} - 1/2 .
\end{align}
\end{subequations}
Here, \(W_{j,l} = W_{j,l}^\dagger = \prod_{i<j} \exp ( - i \pi c_{i,l}^\dagger c_{i,l})\) is the Jordan-Wigner string in a one-dimensional system ensuring the proper anti-commutation relations within each chain, while the operators on different chains commute. The creation and annihilation operators ($c_{j,l}^\dagger$, $c_{j,l}$) therefore satisfy the commutation and anti-commutation relations
 \begin{subequations}\label{eq_stat_c}
     \begin{align}
         [c_{i,l},c_{j,m}^\dagger]&=0 \; \text{ for } \; l\neq m,\\
         \lbrace c_{i,l},c_{j,l}^\dagger\rbrace&=\delta_{ij}.
     \end{align}
 \end{subequations}
The associated particles are therefore transversely bosonic and longitudinally fermionic.

\subsection{Bosonization of the intrachain Hamiltonian}

To understand the intrachain Hamiltonian, we first diagonalize the non-interacting system, and then use Abelian bosonization. 
After the gauge transformation $c_{j,l}\to (-1)^j c_{j,l}$, we apply a Fourier transform,
so that \({H}_l^0 (J_l^z=0, h_l=0) = \sum_k \epsilon_{kl} c_{kl}^\dagger c_{kl} \),  where \(\epsilon_{kl} = -  \sqrt{J_l^2 +  D_l^2} \cos (k a - b_l a)\), \(b_l = \arg (J_l +  i D_l)/a\), and $a$ is the lattice spacing. 
In addition to renormalizing the bandwidth, the DMI therefore shifts the dispersion by  \(b_l \)  in momentum space~\cite{Avalishili2019_LongRange}. 

To bosonize the Hamiltonian~\cite{Giamarchi2003_Quantum}, 
we take the continuum limit and replace lattice site particle creation and annihilation operators with left- and right-moving fermion fields \(L\) and \(R\) according to
\begin{align}
c_{j,l} \,\, \rightarrow \,\, e^{ik_l^R x_j} R_l(x_j) + e^{ik_l^L x_j} L_l(x_j),
\end{align}
where \(k_{l}^{R/L} \equiv \pm k_l + b_l \) are the shifted Fermi momenta.
Here, \(k_l = \pi (1+ 2M_l)/2 a\) is the Fermi momentum of chain $l$, where \(M_l\) is the net $z$-magnetization controlled by the magnetic field $h_l$. Furthermore, $x_j=ja$ and we have set $\hbar =1$. The operators satisfy the commutation relations
\begin{subequations}\label{eq_stat_RL}
     \begin{align}
         [X_{l}(x),Y_{m}^\dagger(x')]&=0 \; \text{ if } \; l\neq m,\\
         \lbrace X_{l}(x),Y_{l}^\dagger(x')\rbrace &=\delta_{XY}\delta(x-x'), \label{eq_stat_RL_2}
     \end{align}
 \end{subequations}
where \(X, Y \in \{L, R\}\). The right- and left-movers can then be expressed in terms of bosonic fields \(\phi_l\) and \(\theta_l\) through \(R_l = e^{-i(\phi_l - \theta_l ) }/\sqrt{2\pi \alpha}\) and \(L_l = e^{i(\phi_l + \theta_l ) }/\sqrt{2\pi \alpha} \), where \(\alpha\) is a short-distance cutoff and we omit the Klein factors. The bosonic fields satisfy the commutation relations 
\begin{subequations}
\begin{align}
[\phi_l(x), \theta_{l'}(x')] &= \delta_{ll'} \frac{i \pi}{2} \operatorname{sgn}(x'- x), \\
[\phi_l(x), \phi_{l'}(x')] &= [\theta_l(x), \theta_{l'}(x')] = 0.
\end{align}
\label{eq_commutationBosonFields}%
\end{subequations}
The creation and annihilation operators \(c_{jl}^\dagger\) and \(c_{jl}\) are then replaced with bosonic fields according to
\begin{align}
c_{j,l} \rightarrow \frac{1}{\sqrt{2\pi\alpha}}\left[T_l^L(x_j) + T_l^R(x_j) \right],
\end{align}
where we have introduced 
\begin{subequations} \label{eq:T_LR}
\begin{align}
T_l^L(x_j) &= e^{i k_l^L x_j} e^{i [ \theta_l(x_j) + \phi_l(x_j) ]}, \\
T_l^R(x_j) &= e^{ i k_l^R x_j} e^{i [\theta_l(x_j ) - \phi_l(x_j) ]} .
\end{align}
\end{subequations}
Within a perturbative renormalization group scheme, the intrachain interactions then give rise to the effective Luttinger liquid (LL) Hamiltonian 
\begin{align}
H_l^0 &=  \int \frac{dx}{2\pi} \, u_l \left[ K_l (\partial_x \theta_l)^2 + \frac{1}{K_l} (\partial_x \phi_l)^2  \right], \label{eq_llHamiltonian}
\end{align}
with chain-dependent renormalized LL parameters \(K_l\) and velocities \(u_l\). These are controlled by the magnetic fields and intralayer out-of-plane exchange interaction $J^z$ and should be computed numerically.

\subsection{Bosonization of the Jordan-Wigner string}

\noindent In terms of density operators in the continuum limit, the Jordan-Wigner string \(W_{j,l}\) takes the form 
\begin{align}
W_{j,l}
\rightarrow \exp \left\{- i\pi \int_{-\infty}^{x_j} dy \left[ \rho_l^0 + \rho_l^R(y) + \rho_l^L(y) \right] \right\},
\end{align}
\noindent where \(\rho_l^0 = k_l / \pi\) is the background particle density and  \(\rho_l^L + \rho_l^R =  L_l^\dagger L_l + R_l^\dagger R_l =  - \nabla \phi_l / \pi\) represents the slowly varying part of the density fluctuations. We have disregarded the off-resonant back-scattering terms. By further disregarding the constant phase factor resulting from the lower limit of the background density integral and the position independent operator \(\phi_l(-\infty)\), we obtain 
\begin{align}\label{eq:T_W}
W_{j,l} \rightarrow   e^{-ik_l x_j + i\phi_l(x_j)} \equiv T_l^W(x_j) .
\end{align}
\noindent Since the operator \(\sum_{i<j} c_{i,l}^\dagger c_{i,l}\) is integer valued, the original Jordan-Wigner string is Hermitian, \(W_{j,l} = W_{j,l}^\dagger\). However, the bosonized form of the Jordan-Wigner string above is not. We therefore symmetrize it and instead use~\cite{Giamarchi2003_Quantum}
\begin{align}
T_{l}^{W,S} =  \frac{1}{2} \left[ T_{l}^W + (T_{l}^W )^\dagger \right] 
\label{eq_suppMat_jwStringSymmetrized}
\end{align}
 as the bosonized form of the Jordan-Wigner string.

\subsection{Interchain interaction}

Introducing the Jordan-Wigner transformation as defined through Eq.~\eqref{eq_jordanWigner}, we obtain the interchain Hamiltonian \(H_\perp = H_\perp^{xy} + H_\perp^z\), with 
\begin{subequations}
\begin{align}
H_\mathrm{\perp}^{xy} &= \frac{J_\perp^{xy}}{2} \sum_{j,l} 
 (c_{j,l}^\dagger c_{j,l+1} + c_{j,l+1}^\dagger c_{j,l} ) W_{j,l} W_{j,l+1},
\\
H_\perp^{z} &=  J_\perp^{z} \sum_{j,l} (c_{j,l}^\dagger c_{j,l} - 1/2) (c_{j,l+1}^\dagger c_{j,l+1} - 1/2),
\end{align}
\end{subequations}
\noindent where we made use of \(W_{j,l} = W_{j,l}^\dagger\). 
Notably, the non-local parts of the Jordan-Wigner strings do not cancel in \(H_\perp^{xy}\) as the product \(W_{j,l} W_{j,l+1}\)  is composed of strings belonging to neighbouring chains. 

Inserting the bosonized representations of the particle creation and annihilation operators according to the usual scheme, we obtain the bosonized interaction  \(H_\perp = H_\perp^{z,0} + H_\perp^{z,1} + H_\perp^{xy}\), with out-of-plane contribution $H_\perp^{z,0}+H_\perp^{z,1}$ and  in-plane contribution $H_\perp^{xy}$ given by 
\begin{widetext}
\begin{subequations}
\begin{align}
H_\perp^{z,0} &=  \quad \,\,\,\, \frac{ a J_\perp^z }{\pi^2} \sum_{l} \int dx \,  \partial_x \phi_l(x) \partial_x \phi_{l+1}(x) ,
\label{eq_hPerp_z0}
\\
H_\perp^{z, 1} &= 
\frac{a J_\perp^z}{(2\pi \alpha)^2} \sum_l \int dx \, 
\big[ (T_{l}^R)^\dagger (T_{l}^L) (T_{l+1}^R)^\dagger ( T_{l+1}^L  )
+ (T_{l}^R)^\dagger  (T_{l}^L) (T_{l+1}^L)^\dagger  (T_{l+1}^R)
+ \mathrm{H.c.} \big], \\
H_\perp^{xy} &= \quad \,\,\, \frac{ J_\perp^{xy}}{2 \pi \alpha} \sum_{l} \int dx \,   \Big\{      
T_{l}^{W,S} T_{l+1}^{W, S}
\Big[ 
 (T_{l}^R)^\dagger (T_{l+1}^R)
+  (T_{l}^L)^\dagger (T_{l+1}^L)  
+ (T_{l}^R)^\dagger (T_{l+1}^L)
 + (T_{l}^L)^\dagger (T_{l+1}^R)
\Big] 
+ \mathrm{H.c.} \Big\},
\end{align}
\label{eq_suppMat_interactionBuildingBlocks}%
\end{subequations}
\end{widetext}
\noindent where $T_l$ depends on the position coordinate $x$ along the chain. We have disregarded the terms that are quadratic in the fermion operators, which can be included in the non-interacting Hamiltonian and simply renormalize the magnetic field terms. The term \(H_\perp^{z,0}\) is bilinear in field gradients and gives an effective kinetic energy term contributing to the quadratic part of the Hamiltonian. The remaining terms represent interactions. 

\subsection{Perturbative generation of interactions}

Using the terms from the interchain exchange interaction as building blocks, we may now construct various symmetry-allowed interactions perturbatively by multiplying together terms. A general form of such an interaction term \(\mathcal{H}_l^\mathrm{int}\) is given by 
\begin{align}\label{eq:inter_1stform}
\mathcal{H}_l^\mathrm{int} =  \int \frac{dx}{4\pi}  g_l  \left[\prod_\delta ( T_{l+\delta}^W )^{\kappa_l^\delta} %(T_{l+\delta}^R)^{s_l^{R,\delta} } (T_{l+\delta}^L)^{s_l^{L,\delta} } 
(T_{l+\delta}^R)^{s_l^{R,\delta} } (T_{l+\delta}^L)^{s_l^{L,\delta} } 
+ \mathrm{H.c.}\right],
\end{align}
\noindent where \(g_l\) is the coupling strength and \(s_{l}^{\alpha, \delta}\) is the net number of particles leaving the branch 
\(\alpha\) (\(L\) or \(R\)) on chain \(l+\delta\) in an interaction centered around chain \(l\). We also introduce the notation \(\sigma^\delta_{l}= s^{R,\delta}_{l} + s_{l}^{L,\delta} \) for the net loss (or gain) of particles in chain \(l + \delta\). From the conservation of particle number, we have \(\sum_\delta \sigma_{l}^\delta =0\).
Furthermore, \(\kappa^{\delta}_{l}\) is an integer such that \(\kappa_{l}^\delta = 0\) when \(\sigma_{l}^\delta  = 0 \mod 2\) and \(\kappa_{l}^\delta = \pm 1\) when \(\sigma_{l}^\delta = 1 \mod 2\), where we used that \((W_{j,l})^2 = 1\). Thus, when \(\sigma_l^\delta\) is odd, \(\kappa_l^\delta\) specifies the direction of the momentum kick due to the Jordan-Wigner string. 
While the two terms given by \(\kappa_l^\delta = \pm 1\) are both generated perturbatively, their respective momentum kicks differ by \(2k_{l+\delta}\), and, thus, only one of the two can typically be resonant for a given set of $\lbrace s_l^{\alpha,\delta} \rbrace$ and system parameters. 

The above interaction represents one of many terms that can be constructed perturbatively. 
We now anticipate that only one interaction term per chain  will be relevant under certain resonance condition (see below),  and that each of these interaction terms can be associated with a particular chain $l$. We thus use the chain index  \(l\) to label the interaction term centered around a given chain $l$. The total interchain interaction Hamiltonian is then
\begin{align}
\mathcal{H}^\mathrm{int} = \sum_l \int \frac{dx}{4\pi} \, g_l  \left(  e^{-i P_l x} e^{i \varphi_l} + \mathrm{H.c.} \right),
\label{eq_interaction_generic}
\end{align}
\noindent where the field \(\varphi_l\) and the total momentum change \(P_l\) are given by
\begin{align}
\varphi_{l} &= \sum_\delta \left[  (s_l^{L,\delta} + s_l^{R,\delta} ) \theta_{l+\delta} +  (s_l^{L,\delta} - s_l^{R,\delta} + \kappa_l^{\delta} ) \phi_{l+\delta} \right],
\\
P_l & =  \sum_\delta [-(s_l^{L,\delta} + s_l^{R,\delta} ) b_{l+\delta} + (s_l^{L,\delta} - s_l^{R,\delta} + \kappa_l^{\delta} )  k_{l+\delta}].
\end{align}
\noindent Since \(s_l^{L,\delta}- s_l^{R,\delta}\) has the same parity as \(\sigma_l^\delta\) and \(\kappa_l^\delta\), the integer \( 2 n_l^\delta \equiv s_l^{L,\delta} - s_l^{R,\delta} + \kappa_l^\delta\) is always even. 
We then write
\begin{align}
\varphi_l &= \sum_\delta \left( \sigma_l^\delta \theta_{l+\delta} + 2n_l^{\delta} \phi_{l+\delta} \right),  \label{eq_bosonFieldForm} \\
P_l &= \sum_\delta \left(
- \sigma_l^\delta b_{l+\delta} + 2n_l^\delta k_{l+\delta}
\right) .
\label{eq_momentumPl}
\end{align}
The interaction term in Eq.~\eqref{eq_interaction_generic} is finite  when the {\it resonance condition} $P_l=2\pi p_l/a$ is satisfied, where $p_l \in \mathbb{Z}$.

\begin{figure*}
    \centering
    \includegraphics[width=0.94\textwidth]{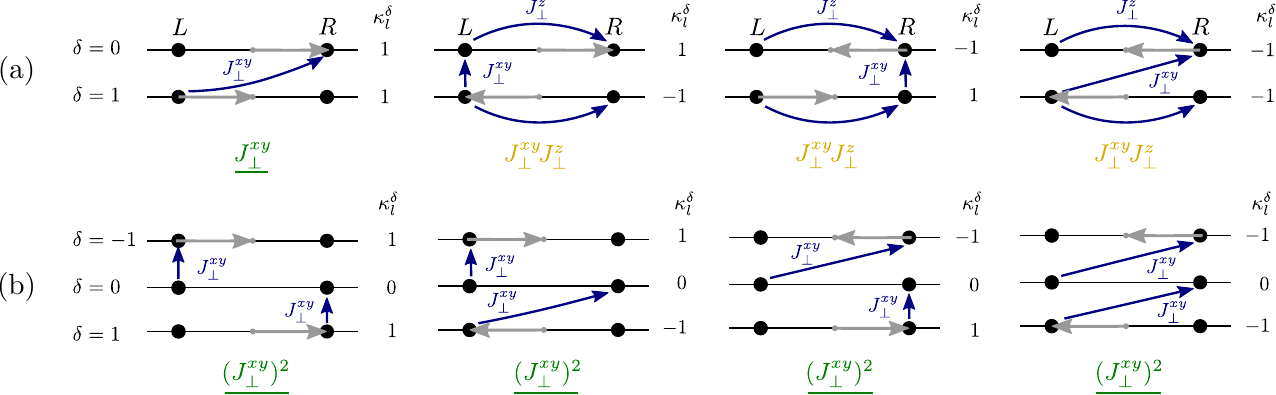}
    \caption{ Scattering processes that contribute to the interaction terms defined by \(\{\sigma_l^\delta, n_l^\delta\}\). Horizontal black lines represent chains,  black dots are the Fermi points on the respective chains, blue arrows are particle scattering processes, and grey arrows are the associated momentum kicks due to the Jordan-Wigner strings. For chains where the net particle number changes, there are two options (\(\kappa_{l}^\delta  = \pm 1\)) for the direction of the momentum kick due to the Jordan-Wigner strings. Thus, a given bosonic interaction term (specified by $\{\sigma_l^\delta, n_l^\delta\}$) can arise through different scattering processes, which in general give different contributions to the effective coupling strength $g_l$ associated with the bosonic interaction term. Further processes can be constructed from the diagrams above by replacing single arrows with a chain of arrows starting and ending at the same two points. The order of %magnitude of 
    the interaction amplitude associated with a particle scattering process is indicated below the diagram, with leading order amplitudes in green, and subleading amplitudes in yellow. Since for a given construction, all these processes correspond to the same choice of \(\{\sigma_l^\delta, n_l^\delta\}\), they are all resonant (or off-resonant) at the same time. 
    %\et{\sout{For resonant processes, the total momentum change is an integer multiple of $2\pi$. }} 
    (a) First-level FSQHE construction with \((\sigma,n) = (1, 1)\).  The processes are of different order in the exchange coupling amplitudes, and the leftmost process is expected to dominate.  
    (b) Second-level FSQHE construction with \((\sigma, n_0, n_1) = (1, 1, 0)\). All four processes are of the same order in the exchange coupling amplitudes. 
    Although not shown in the figure, note that the locations of the left and right shifted Fermi momenta vary from chain to chain. 
    }
    \label{fig_suppMat_scatteringProcesses}
\end{figure*}

In the following, we refer to terms of the form in Eq.~\eqref{eq_interaction_generic} as \textit{bosonic interaction terms}. 
These bosonic interaction terms can be constructed by combining interchain interaction terms from Eq.  \eqref{eq_suppMat_interactionBuildingBlocks}. 
For a given bosonic interaction term, we expect many possible combinations, which all contribute to a common coupling strength for the bosonic interaction term. 
The integers $\{\sigma_l^\delta, n_l^\delta\}$ fully characterize each effective bosonic interaction and are related to the integers $\{s_l^{L,\delta}, s_l^{R,\delta}\}$ through
\begin{subequations}\label{eq:s}
\begin{align}
s_l^{L, \delta}  &= \frac{1}{2} (\sigma_l^\delta + 2 n_l^\delta - \kappa_l^\delta), \\
s_{l}^{R,\delta} &= \frac{1}{2} (\sigma_l^\delta - 2n_l^\delta + \kappa_l^\delta) .
\end{align}
\label{eq_sParameters}%
\end{subequations}
When \(\sigma_l^\delta\) is even, $\kappa_l^{\delta}=0$, so that fixing \(\sigma_l^\delta\) and \(n_l^\delta\) uniquely fixes \(s_l^{L,\delta}\) and \(s_l^{R,\delta}\). When \(\sigma_l^\delta\) is odd, there are two possibilities for \(\kappa_l^\delta\), 
and Eq.~\eqref{eq:s} gives two different possibilities for \(s_l^{L,\delta}\) and \(s_l^{R,\delta}\). Therefore, given a bosonic interaction, i.e. a set of integers \(\lbrace \sigma_l^\delta, n_l^\delta \rbrace\) for all $\delta$, there are $2^{n_o}$ possibilities for the set \(\lbrace s_l^{L,\delta},s_l^{R,\delta}\rbrace\), where $n_o$ is the number of odd integers \( \sigma_l^\delta\) in the bosonic interaction. 
These possibilities represent different scattering processes, 
and in general, their amplitudes are of different order in the exchange coupling strengths. 
We provide explicit examples of this in Fig.~\ref{fig_suppMat_scatteringProcesses},
as discussed in Secs.~\ref{sec_fsqhe} and \ref{sec_higherLevel} below.

Above, we considered very general interactions. In the following, we construct phases where the interactions specified by the integers \(\lbrace \sigma_l^\delta, n_l^\delta \rbrace\)  in Table~\ref{tab_integerDefinitions} dominate the physics. 

\begin{table}[!btp]
    \caption{
    Notation for the integers \(\sigma_l^\delta\) and \(n_l^\delta\) in the interactions giving rise to the first- and second-level FSQHE phases.}
    \centering
    \begin{tabular}{llll}
    \hline
     & \hspace{1.3cm} & $\sigma_l^\delta$ \hspace{0.2cm} & $n_l^\delta$  \\
    \hline
    1st-level FSQHE \hspace{0.3cm}   & $\delta = \,\,\; 0$ & $-\sigma$ & $n$   \\
    & $\delta = +1$     & $+\sigma$ & $n$  \\
    \hline 
     &  $\delta=-1$ & $-\sigma$ & $n_1$ \\
    2nd-level FSQHE \hspace{0.3cm} & $\delta = \,\,\; 0$ & \,\, $0$ & $n_0$ \\
    & $\delta = +1$ & $+\sigma$ & $n_1$ \\
    \hline
    \end{tabular}
    \label{tab_integerDefinitions}
\end{table}

\section{Fractional spin quantum Hall effect}\label{sec_fsqhe}
To realize a FSQHE (or CSL) phase with gapless chiral edge modes, we need to break the
$P$ and $T$ symmetries.  
The simplest bosonic interactions breaking the symmetries are specified by \(\sigma_l^1 = - \sigma_l^0 \equiv \sigma  \), \(n_l^1 = n_l^0 \equiv n\), and \(n_l^\delta = \sigma_l^\delta = 0\) for any other \(\delta\), as listed in Table~\ref{tab_integerDefinitions}. 
Here, we introduced integers \(\lbrace \sigma_l^\delta, n_l^\delta \rbrace\) characterizing the first-level FSQHE interaction.

Physical insight into the scattering processes producing the bosonic interaction specified by \(\lbrace \sigma_l^\delta, n_l^\delta \rbrace\)  is obtained by determining \(s_l^{\alpha, \delta}\) through Eq.~\eqref{eq_sParameters}. 
As an example, we consider \((\sigma, n) = (1, 1)\). Choosing \(\kappa_l^0 = \kappa_l^1 = 1\), we obtain \((s_{l}^{L,0}, s_{l}^{R,0}, s_{l}^{L,1}, s_l^{R,1}) = (0, -1, 1, 0)\). The corresponding scattering process is represented as a single particle scattering from the left branch of chain \(l+1\) to the right branch of chain \(l\), as indicated in the leftmost process of Fig.~\ref{fig_suppMat_scatteringProcesses}(a). 
By also calculating the integers \(s_l^{L,\delta}\) and \( s_l^{R,\delta}\) corresponding to the other choices of \(\kappa_l^{\delta}= \pm 1\), we obtain the remaining three processes in Fig.~\ref{fig_suppMat_scatteringProcesses}(a).
Among the four possibilities, the leftmost process is the simplest and is generated by a term in Eq.~\eqref{eq:hperp} already at first order in the interaction. We, therefore, expect its contribution to determine the sign and magnitude of $g_l$ and focus on \(g_l \propto J_\perp^{xy}\).  
For generic $\sigma$ and $n$, the lowest order contribution is of order \((J_\perp^{xy} )^\sigma (J_\perp^z)^{\gamma}\), with exponent \(\gamma= \operatorname{max}\{0, |n|- (|\sigma|+1)/{2}\}\) when \(\sigma\) is odd and \(\gamma = \operatorname{max}\{0, |n|- |\sigma|/2\}\) when \(\sigma\) is even.

The bosonic interaction term introduced above opens a gap in the bulk spectrum (Sec.~\ref{sec_fsqhe_bulkGap}) when it is resonant (Sec.~\ref{sec_fsqhe_choice}) and relevant (Sec.~\ref{sec_fsqhe_scalingDim}), so that it can dominate the physics to produce the FSQHE phase. For the simplest interaction with \((\sigma,n) = (1, 1)\), this can be achieved by choosing the magnetization and dispersion relation shift profiles shown in Fig.~\ref{fig_parameterChoice}. We also provide an example of how this can be done for the phase with \((\sigma,n) = (1,2)\). Subsequently, we discuss the properties of the FSQHE phase and calculate the fractional spin of the elementary spin excitations (Sec.~\ref{sec_fsqhe_fracSpin}) as well as the two-terminal spin conductance (Sec.~\ref{sec_fsqhe_conductance}), which is the analogue of the fractional Hall conductance in an electronic system. Finally, we argue explicitly that the phases satisfy all the necessary criteria to be called CSLs (Sec.~\ref{sec_fsqhe_csl}).

\subsection{Bulk gap}\label{sec_fsqhe_bulkGap}

\begin{figure}
    \centering
    \includegraphics[width=0.95\columnwidth]{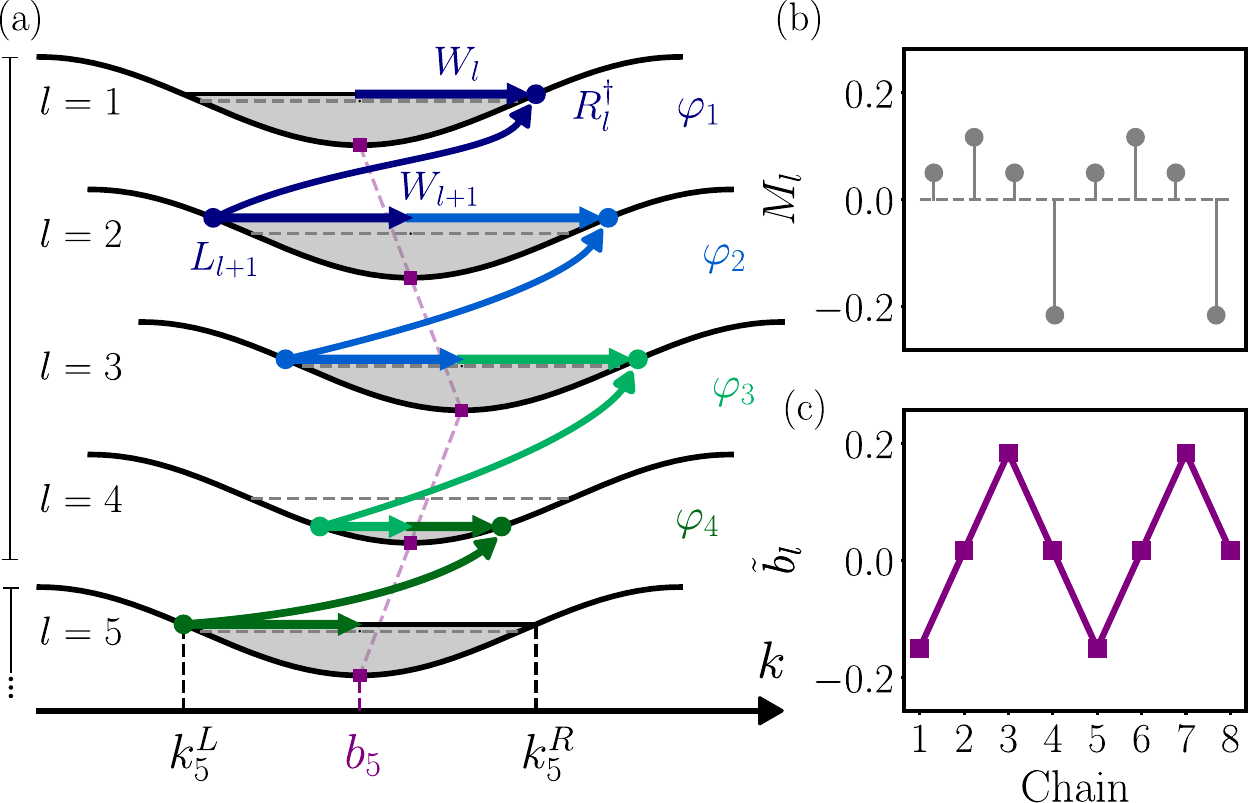}
    \caption{Construction of FSQHE in a system formed by a repeating unit cell consisting of four spin chains. (a) Dispersion relation of the Jordan-Wigner particles in the chains. The particle filling fractions  \(k_l/\pi = 1/2 + M_l\) are indicated by a solid line, where \(M_l\) is the magnetization of chain $l$. Half-filling is indicated with a dashed line. The dispersion relation of each chain is shifted by $b_l$ due to the chain-dependent intrachain DMI.  The simplest particle scattering process corresponding to the interaction associated with the field \(\varphi_l\) is illustrated with an arrow. 
    When the net change in the number of particles in a chain is odd, 
    the Jordan-Wigner string \(W_{l}\) gives rise to an additional momentum kick \(\pm k_l\), as indicated by the horizontal arrows on the Fermi surfaces. The momenta associated with a given interaction add up to a multiple of $2\pi$ (resonance condition). In the FSQHE phase, gaps open at the Fermi points involved in the interactions, leaving two chiral gapless modes on the edges. Profiles (b) of magnetization $M_l$ and (c) of the corresponding shift \(b_l\) needed for the FSQHE construction. \\
    }
    \label{fig_parameterChoice}
\end{figure}

When the resonance condition $P_l = 2\pi p_l/a$ is satisfied, the interaction Hamiltonian is 
\begin{align}
\label{eq_fqheInteraction}
\mathcal{H}^{\mathrm{res}} = \sum_{l=1}^{N-1} \int \frac{dx}{2\pi} \, g_l 
\cos \varphi_l
\end{align}
with commuting boson fields \(\varphi_l\) given by
\begin{align}\label{eq_fsqhe_varphiDef}
\varphi_l = \sigma ( \theta_{l+1}- \theta_l) + 2 n (\phi_l + \phi_{l+1}),
\end{align}
where $l = 1, ..., N-1$. % $l \in [1, N-1]$. 
Introducing the chiral fields \(\eta_l^L\) and \(\eta_l^R\) defined by 
\begin{subequations}
\begin{align}
\eta_l^L &= \sigma \theta_l + 2n \phi_l, \\
\eta_l^R &= \sigma \theta_l - 2n \phi_l,
\end{align}
\end{subequations}
we have \(\varphi_l  = \eta_{l+1}^L - \eta_{l}^R\).
In the strong coupling limit, the interaction pins the $(N-1)$ commuting bosonic fields $\varphi_l$, gapping \(2(N-1)\) bosonic modes. As seen in Fig.~\ref{fig_parameterChoice}, this leaves the two gapless chiral edge modes \(\eta_1^L\) and \(\eta_N^R\), and gives rise to a FSQHE phase.

\subsection{Choice of magnetizations and DMI strengths}\label{sec_fsqhe_choice}

In this section, we specify a choice of magnetization and DMI strength patterns that can be used to obtain the first-level FSQHE phase. We then argue that no other interactions satisfying the resonance condition have a smaller scaling dimension. While many different choices of magnetization and DMI strength can make a given target interaction resonant, our goal is simply to show that it is possible.
%demonstrate that at least one is possible. 

For \(n_l^{1} = n_l^0 = n\) and \(\sigma_l^1 = - \sigma_l^0 \equiv \sigma \), the resonance condition assumes the form
\begin{align}
P_l \equiv 2n (k_l + k_{l+1} ) + \sigma(b_l - b_{l+1}) = 2\pi p_l/a,
\end{align}
\noindent with \(p_l \in \mathbb{Z}\). We now introduce the filling fractions \(f_l = k_l a/\pi \equiv 1/2 + M_l \in [0, 1] \) as well as the rescaled dimensionless dispersion relation shifts \(\tilde{b}_l = b_l a/2\pi\). In the following, we consider a construction with a repeating basis of four chains, so that every fourth chain has the same DMI strengths and fillings. The resonance condition then takes the form 
\begin{align}
\label{eq_matrixResonanceCondition}
A \bm{f} + D \tilde{\bm{b}} = \bm{p},
\end{align}
where the \(4\times 4\)-matrices \(A\) and \(D\) are given by
\begin{align}
A = n\begin{pmatrix} 1 & 1 & 0 & 0 \\ 0 & 1 & 1 & 0\\ 0 & 0 & 1 & 1 \\ 1 & 0 & 0 & 1 \end{pmatrix},
\quad 
D =  \sigma \begin{pmatrix} 1 & -1 & 0 & 0 \\ 0 & 1 & -1 & 0\\ 0 & 0 & 1 & -1 \\ -1 & 0 & 0 & 1 \end{pmatrix},
\end{align}
and \(\tilde{\bm{b}}\) and \(\bm{p}\) are the column vectors with \(\tilde{b}_l\) and \(p_l\) as elements. 
Adding up the four equations comprising the resonance condition, we have \(\sum_l f_l/4 = \sum_l p_l/ 8 n\). Thus, it is already clear that the resonance condition can only have solutions for average fillings that are integer multiples of \(1/8n\)~\footnote{This is similar to the standard quantum Hall effect, where the resonance condition is also set by the filling factor $\nu$ and not by the magnetic field and the electron density independently.}. 

The above set of four equations in the eight variables \(f_l\) and \(\tilde{b}_l\) is underdetermined. 
The matrix \(\begin{pmatrix} A & D\end{pmatrix}\) of size \(4 \times 8\) has rank four, and therefore a nullspace of dimension four.  The general solution is
\begin{align}
\begin{pmatrix} \bm{f} \\ \tilde{\bm{b}} \end{pmatrix} = 
%\frac{1}{2} \begin{pmatrix} 1 \\ 1 \\ 1 \\ 1 \\ 0 \\ 0\\ 0 \\ 0 \end{pmatrix}
\begin{pmatrix} \bm{f}_0 \\ \tilde{\bm{b}}_0 \end{pmatrix}
+ \begin{pmatrix} 
0 & +1 & +4 \sigma & +3 \sigma \\ 
0 & -1 & +3 \sigma & -4 \sigma \\ 
0 & +1 & -4 \sigma & -3 \sigma \\ 
0 & -1 & -3\sigma  & +4 \sigma \\ 
1 & 0 & -3n & +4n \\ 
1 & 0 & +4n & +3n \\ 
1 & 0 & +3n & -4n \\ 
1 & 0 & -4n & -3n \end{pmatrix}
\begin{pmatrix} \beta_0 \\ \beta_1 \\ \beta_2 \\ \beta_3 \end{pmatrix},
\end{align}
where \(\begin{pmatrix} \bm{f}_0^T, & \tilde{\bm{b}}_0^T \end{pmatrix}^T\) is a particular solution, the column vectors of the right-hand side matrix span the nullspace of \(\begin{pmatrix} A & D \end{pmatrix}\), and \(\{\beta_i\}\) are some coefficients. 

For simplicity, we look for solutions with \(p_l = n\) in the following, corresponding to average filling \(1/2\). 
Clearly, \(\begin{pmatrix} \bm{f}_0^T, & \tilde{\bm{b}}_0^T \end{pmatrix} = \begin{pmatrix} 1, & 1, & 1, & 1, & 0, & 0, & 0, & 0 \end{pmatrix}/2\) then solves the resonance condition and can be used to generate any other solution. 
A particular non-trivial example is
\begin{subequations}\label{eq:ex_sol_1}
    \begin{align}
\tilde{\bm{b}} &= \frac{1}{62 \sigma} \begin{pmatrix} -9, & 1, & 11, & 1 \end{pmatrix} ^T, \\
\bm{M} &= \frac{1}{62 n} \begin{pmatrix} 3, & 7, & 3, & -13 \end{pmatrix}^T.
\label{eq_suppMat_1stMagnetizations}
\end{align}
\end{subequations}
The corresponding profiles for $\tilde{\bm{b}}$ and $\bm{M}$ with $(\sigma, n) = (1,1)$ are shown in Fig.~\ref{fig_parameterChoice}.

While satisfying the resonance condition is necessary to make the target interaction dominate the physics, it is not sufficient. The system parameters \(\tilde{\bm{b}}\) and \(\bm{M}\) above have also been chosen to satisfy other requirements. 
First, since \(M_l\) represent magnetizations (of spin-1/2 per site), they need to satisfy \(|M_l| < 1/2\). This is clearly the case in Eq.~\eqref{eq_suppMat_1stMagnetizations}. 
Second, the target bosonic interaction needs to be more relevant than any other bosonic interaction
which also satisfies the resonance condition for the given choice of \(M_l\) and \(\tilde{b}_l\).
In the following, we restrict our analysis to interactions involving three neighbouring chains at most (\(|\delta|\leq 1\)), as we expect the amplitudes of other interactions to be too weak to compete, even if they are resonant. 
In Eq.~\eqref{eq:ex_sol_1},  \(\tilde{b}_l\) and \(M_l\) are rational numbers. Therefore, the resonance condition $P_l=2\pi p_l/a$ 
can be rewritten as linear and homogeneous Diophantine equations for the integers \(\sigma_l^\delta,\ n_l^\delta\), and \(p_l\). By construction, \(\sigma_l^1 = -\sigma_l^{0} = \sigma\) and \(n_l^0 = n_l^1 = n\) satisfy these Diophantine equations. We then make a list of other bosonic interactions that also satisfy the Diophantine equations. Finally, we check if any of these are more relevant than the target interaction. For this, we need a precise notion of relevance and, here, we will use the weak coupling scaling dimension of the bosonic interaction, as discussed in further detail in Sec.~\ref{sec_fsqhe_scalingDim} below~\footnote{To be more precise, we use the weak coupling scaling dimension obtained for \(J_\perp^z = 0\). Furthermore, we calculate the scaling dimension by assuming \(K_l = K\). When comparing the scaling dimension of the target interaction with the scaling dimension of a competing interaction, we use the value of \(K\) which minimizes the scaling dimension of the target interaction.}. For the choices of \(M_l\) and \(\tilde{b}_l\) in Eq.~\eqref{eq:ex_sol_1}, we find that none of the other resonant interactions are more relevant than the target interaction for both \((\sigma, n) = (1, 1)\) and \((\sigma, n) = (1, 2)\). Thus, if the target interaction flows to strong coupling, it dominates the physics.

\subsection{Scaling dimension and phase diagram}\label{sec_fsqhe_scalingDim}
The FSQHE is realized in the strong coupling regime when all the bare coupling strengths \(g_l\) are large. Alternatively, the strong coupling regime is also reached for weak transverse coupling strengths ($J_\perp \approx 0$) when the weak coupling scaling dimensions \(\Delta_l\) of the interactions are smaller than 2. Then, the interaction in Eq.~\eqref{eq_fqheInteraction} is relevant and dominates the low-energy physics. 

The scaling dimension \(\Delta_l\) of the interaction term \(\mathcal{I}_l = \exp \{i [\sigma(\theta_{l+1} - \theta_{l}) + 2n(\phi_{l+1} + \phi_{l})  ] \} \) occurring in Eq.~\eqref{eq_fqheInteraction} can be determined from the decay of the correlation function 
\(\langle \mathcal{I}_l^\dagger(x) \mathcal{I}_l(0)\rangle \propto 1/x^{2\Delta_l} \)~\footnote{Alternatively, one may extract the scaling dimension by deriving the renormalization group flow of the coupling constants to linear order in the interchain coupling strength \(J_\perp\). The result is exactly the same.}. 
Neglecting the bi-linear kinetic term in Eq.~\eqref{eq_hPerp_z0} for now, the intrachain scaling dimension calculated solely with the LL Hamiltonian in Eq.~\eqref{eq_llHamiltonian} is 
\begin{align}\label{eq_fsqhe_scalingDim_bare}
\Delta_l|_{J_\perp^z = 0} =   n^2 \left(K_l+ K_{l+1} \right) + \frac{\sigma^2}{4} \left(\frac{1}{K_l}+ \frac{1}{K_{l+1}}\right).
\end{align}
Thus, when \((\sigma, n)=(1,1)\), the interaction \(\mathcal{H}^\mathrm{res}\) is marginal when \(K_l=K_{l+1}=1/2\) and irrelevant for \(K_l \neq 1/2\)~\cite{Ronetti2022_Fractional}. 
However, the interchain interaction ($J^z_\perp>0$) can make the first-level FSQHE interaction relevant even at weak coupling as we show below. 

We calculate the scaling dimension of the field \(\mathcal{I}_l(x)\) when the quadratic part of the Hamiltonian originates from the intrachain LL Hamiltonian in Eq.~\eqref{eq_llHamiltonian} and the bilinear interchain interaction in Eq.~\eqref{eq_hPerp_z0}, so that \(H^\mathrm{quad} = H^\mathrm{quad}_0 + H^\mathrm{quad}_\perp\), with 
\begin{subequations}
\label{eq_fsqhe_scalingDim_Hquad}
\begin{align}
H^\mathrm{quad}_0 &= \sum_{l} \int \frac{dx}{2\pi}    \left[u_l  K_l (\partial_x \theta_l)^2   
+ \frac{u_l}{K_l} (\partial_x \phi_l)^2 \right],  \\  
H^\mathrm{quad}_\perp &= \sum_{lm} \int \frac{dx}{2\pi}  u (\partial_x \phi_l) v_{lm} (\partial_x \phi_m ).
\end{align}
\end{subequations}
\noindent Here, \(u\) is a velocity scale, and we set it to the geometric mean \(u = \left( \prod_l u_l\right)^{1/N}\). The dimensionless coupling coefficient $v_{lm}$ describes the kinetic coupling between different chains. Direct comparison with Eq.~\eqref{eq_hPerp_z0} gives \(v_{lm} = (v_\perp^z/2) (\delta_{l,m+1}+ \delta_{l,m-1})\), where the dimensionless interchain out-of-plane interaction strength is \(v_\perp^z = 2 a J_\perp^z/\pi u\).

To diagonalize a Hamiltonian of this form, we introduce transformed fields \(\tilde{\phi}_n\) and \(\tilde{\theta}_n\) through \({\phi}_n = \sum_m T_{nm}^\phi \tilde{\phi}_m\) and \( {\theta}_n = \sum_m T_{nm}^\theta \tilde{\theta}_m\), where the transformation matrices $T$ satisfy \((T^\phi)^T T^\theta = (T^\theta)^T T^\phi = 1\) to preserve the commutators~\cite{Yurkevich2013_DualityMultichannelLuttingerb}. 
We choose \(T\) so that the \(\phi\)-sector is diagonalized, while the \(\theta\)-sector remains diagonal. This is achieved through \(T_{nm}^\theta = \Gamma_n U_{nm}\) and \(T_{nm}^\phi = \Gamma_n^{-1} U_{nm}\), 
where \(\Gamma_n = (u_n K_n /u)^{-1/2}\) and \(U\) is the orthogonal matrix that diagonalizes \(W_{nm} = (u_n/u)^2 \delta_{nm} + \Gamma_{n}^{-1} v_{nm} \Gamma_m^{-1} \), so that \((U^T W U)_{nm} = d_n \delta_{nm}  \). In terms of the new fields, the Hamiltonian takes the familiar Luttinger liquid form with renormalized velocity parameter \(\tilde{u}_m = u \sqrt{d_m}\) and LL parameter \(\kappa_m = 1/\sqrt{d_m}\). This immediately allows us to write down the correlation functions for the fields \(\tilde{\theta}_n\) and \(\tilde{\phi}_n\). In turn, this determines the correlation functions for \(\theta_n\) and  \(\phi_n\) as well as the scaling dimensions of the interaction terms. 

In the following, we consider periodic boundary conditions and let \(K_l = K\) and \(u_l = u\) for all $l$. 
Although this may not be realistic for the FSQHE construction with chain-dependent magnetizations, it suffices to illustrate the general mechanism of out-of-plane interchain interactions making the FSQHE interaction with \((\sigma, n) = (1, 1)\) relevant even at weak coupling. 

\begin{figure}
    \centering
    \includegraphics[width=1\columnwidth]{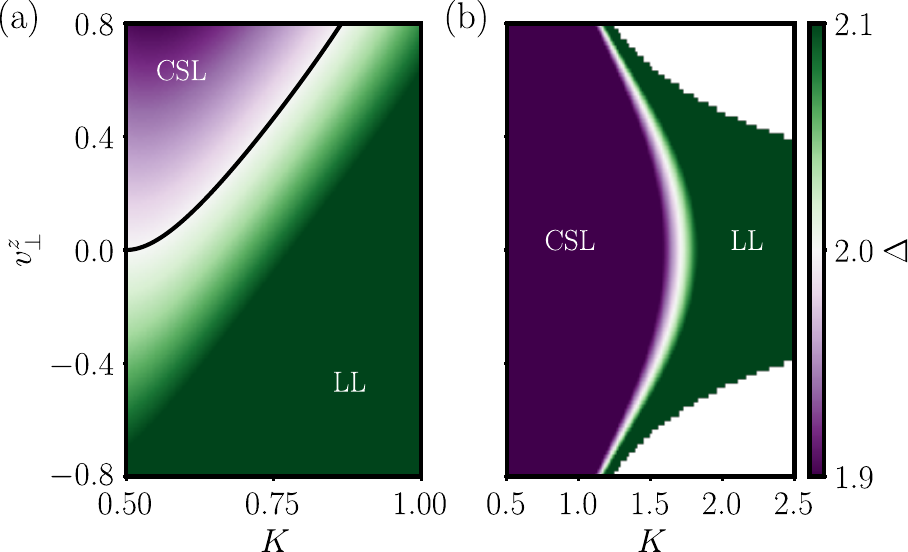}
    \caption{
    Weak coupling phase diagrams with CSL (or FSQHE) and LL phases as deduced from the condition \(\Delta < 2\) on the scaling dimension.
    (a) FSQHE construction with \((\sigma, n) = (1, 1)\), giving rise to a phase with $s^*=1/2$ and one chiral edge state. In the absence of bilinear interchain interaction proportional to \(v_\perp^z\), the interaction giving rise to the FSQHE phase is  
    marginal when \(K = 1/2\), and otherwise irrelevant. 
    The bilinear interchain interaction renormalizes the scaling dimension so that it can become relevant also at weak coupling.  (b) Second-level construction  
    with \( (\sigma, n_0, n_1) = (1, 1, 0)\). The phase hosts two chiral edge states and elementary spin excitations with spin $s^* = 1$.
    The interaction can be relevant even in the absence of the bilinear interchain coupling. } 
    \label{fig_phaseDiagram}
\end{figure}

We now calculate the scaling dimension \(\Delta^\mathrm{1st}\) of $\mathcal{I}_l(x)$ 
in the limit \(N \rightarrow \infty\) as shown in Appendix~\ref{app_scalingDimension}. We find \(\Delta^\mathrm{1st} = \Delta_\theta^\mathrm{1st} + \Delta_\phi^\mathrm{1st}\), where the respective contributions from the correlation functions of \(\theta\) and \(\phi\) are given by 
\begin{subequations}
\begin{align}
\Delta_\theta^\mathrm{1st} &= \,\,\,\, \frac{\sigma^2}{2K} [ J_0^+(v_\perp^z K) - J_1^+(v_\perp^z K)] ,\\
\Delta_\phi^\mathrm{1st} &= 2n^2 K [J_0^-(v_\perp^z K) + J_1^-(v_\perp^z K)] ,
\end{align}
\end{subequations}
where \(J_n^\pm(a)\) is the function
\begin{align}
J_n^\pm ( a) &= \int_0^{2\pi} \frac{dp}{2\pi} (1 + a\cos p)^{\pm 1/2} \cos^n p \,  .
\end{align}
\noindent As expected from the known result for \(v_\perp^z=0\) in Eq.~\eqref{eq_fsqhe_scalingDim_bare}, we have \(J_0^+(0) - J_1^+(0) = J_0^-(0) + J_1^-(0) = 1\). The above expressions can also be expressed in terms of complete elliptic integrals of the first and second kind~\cite{AbramowitzStegun1965}, as given in Appendix~\ref{app_scalingDimension}. From the relevance condition \(\Delta^\mathrm{1st} < 2\), we may now deduce the weak coupling phase diagram. As shown in Fig.~\ref{fig_phaseDiagram}(a), there exists a parameter regime where the first-level FSQHE interaction is relevant even at weak coupling.

\subsection{Fractional spin}\label{sec_fsqhe_fracSpin}

In the strong coupling limit, the bulk modes are gapped.
The elementary massive excitations are kinks, i.e. single phase slips of \(2\pi\) in one field $\varphi_l$. The effective spin value \(s^*\) along the \(\hat{z}\)-direction  of these excitations can be calculated through 
\begin{align}
s^* =
- \frac{1}{\pi} \sum_{l=1}^{N} \int dx \; \partial_x \phi_l(x) ,
\label{sec_fracSpin_def}
\end{align}
\noindent where \(\{\phi_l(x)\}\) represents a field configuration hosting a single kink~\cite{Ronetti2022_Fractional}. Expressing the fields \(\{\phi_l\}\) in terms of the boson fields \(\{\varphi_l\}\) defined in Eq.~\eqref{eq_fsqhe_varphiDef}, we obtain 
\begin{align}
s^* =
%\frac{1}{2n} \int \frac{dx}{2\pi} \;  \left[ \partial_x \varphi_\mathrm{edge}(x)  +   \sum_{l=1}^{N-1}\partial_x \varphi_l(x) \right], \label{eq:frac_spin_calc}
\frac{1}{2n}\sum_{l=1}^{N-1}  \int \frac{dx}{2\pi} \;  \partial_x \varphi_l(x), \label{eq:frac_spin_calc}
\end{align}
where we have disregarded the edge contribution arising from the field derivative  \(\partial_x (\eta_1^L-\eta_N^R)\).
In the presence of a single kink, the field \(\varphi_{l^*}\) associated with chain \(l^*\) jumps by \(2\pi\). The spin of the elementary excitation is therefore
\begin{align}\label{eq:frac_spinRes}
s^* = \frac{1}{2n}, 
\end{align}
\noindent which takes on, in general, {\it fractional values}. 

\subsection{Spin conductance}\label{sec_fsqhe_conductance}

\noindent In this section, we calculate the two-terminal spin conductance in the first-level FSQHE phase following Ref.~\cite{Oreg2014_Fractional}. In a two-terminal setup, this is precisely the Hall spin conductance of the system. An alternative method to calculate the spin conductance is given in Ref.~\cite{Ronetti2022_Fractional}.

We consider a finite-size system of length \(L\) and attach leads at \(x=\pm L/2\) so that spin currents, transporting $z$-magnetization through the system, can flow ~\cite{Meier2003_MagnetizationTransport, Nakata2017}.
The incoming and outgoing magnetization currents \(I_l^\alpha\) and \(O_l^\alpha\) associated with chain \(l\) and branch \(\alpha \in \{L\leftrightarrow -1, R\leftrightarrow +1 \}\) are given by
\begin{subequations} \label{eq:currents}
\begin{align}
I_l^\alpha &= \beta \partial_t \phi_l^\alpha |_{x=+x_l^\alpha},  \\
O_l^\alpha &= \beta \partial_t \phi_l^\alpha |_{x=-x_l^\alpha} ,
\end{align}
\end{subequations}
\noindent where \(\phi_l^\alpha = \theta_l - \alpha \phi_l\), and \(x_l^\alpha = - \alpha L/2\). 
Thus, the right-moving branch corresponds to an incoming current on the left of the system and an outgoing current on the right of the system. The constant \(\beta =  g \hbar \mu_B / \pi\) follows from the definition of the magnetization current, and can be derived from the spin continuity equation~\footnote{From the Jordan-Wigner transformation, the magnetization density is given by \( M^z = M_0^z - (\hbar g \mu_B / \pi) \partial_x \phi(x)  \), where \(M_0^z\) is the background magnetization and the second term represents fluctuations. Thus, inserting into the continuity equation for magnetization, \(\partial_t M^z + \partial_x j_M = 0\), we can identity the magnetization current \(j_M = (\hbar g \mu_B /\pi) \partial_x \phi\). }. Here, \(g\) is the \(g\)-factor and \(\mu_B\) the Bohr magneton.

To determine the conductance at stationarity, we need to know the $2N$ outgoing currents in terms of the $2N$ incoming currents. In the strongly coupled regime and at low energy, 
\(\varphi_l = \eta_{l+1}^L - \eta_l^R \) is fixed to a constant, and \(\partial_t \varphi_l = 0\). 
Enforcing this at \(x=\pm L/2\) and using the relations
\begin{align}
\begin{pmatrix} \eta_l^L \\ \eta_l^R \end{pmatrix}
= \begin{pmatrix} \sigma/2 + n & \sigma/2 - n \\ \sigma/2 - n & \sigma/2 + n \end{pmatrix}
\begin{pmatrix} \phi_l^L \\ \phi_l^R \end{pmatrix},
\end{align}
\noindent we obtain the \(2N-2\) equations 
\begin{subequations}
\begin{align}
\left( \frac{\sigma}{2} - n \right) O_l^L + \left( \frac{\sigma}{2} + n \right) I_l^R = \left( \frac{\sigma}{2} + n \right) O_{l+1}^L + \left( \frac{\sigma}{2} - n \right) I_{l+1}^R ,
\\
\left( \frac{\sigma}{2} - n \right) I_l^L + \left( \frac{\sigma}{2} + n \right) O_l^R = \left( \frac{\sigma}{2} + n \right) I_{l+1}^L + \left( \frac{\sigma}{2} - n \right) O_{l+1}^R .
\end{align}
\end{subequations}
The final two equations necessary to uniquely determine the outgoing currents are obtained from the free propagation of the edge modes, which implies that \( \partial_t \eta_1^L(-L/2) = \partial_t \eta_{1}^L(L/2)\) and \(\partial_t\eta_N^R(-L/2) = \partial_t \eta_N^R(L/2)\). This gives the equations 
\begin{subequations}
\begin{align}
\left( \frac{\sigma}{2} + n \right) O_1^L + \left( \frac{\sigma}{2} - n \right) I_1^R &= \left( \frac{\sigma}{2} + n \right) I_1^L + \left( \frac{\sigma}{2} - n \right) O_1^R,
\\
\left( \frac{\sigma}{2} - n \right) O_N^L + \left( \frac{\sigma}{2} + n \right) I_N^R &= \left( \frac{\sigma}{2} - n \right) I_N^L + \left( \frac{\sigma}{2} + n \right) O_N^R.
\end{align}
\end{subequations}
\noindent The relation between the incoming and outgoing currents then takes the form \(A \bm{O}= B \bm{I}\), where 
\begin{subequations}
\begin{align}
\bm{I} &= \begin{pmatrix} I_1^L, & I_1^R, & \cdots & I_N^L, & I_N^R \end{pmatrix}^T, \\
\bm{O} &= \begin{pmatrix} O_1^L, & O_1^R, & \cdots & O_N^L, & O_N^R \end{pmatrix}^T,
\end{align}
\end{subequations}
and \(A\) and \(B\) are the \(2N \times 2N\)-matrices specified in Appendix~\ref{app_conductance}. From this, we obtain \(\bm{O} = S \bm{I}\) with \(S = A^{-1} B\). We now subject the system to an energy bias \(V = g \mu_B \Delta B\), where \(\Delta B\) is the difference in the magnetic field of the left and right leads. Assuming there is an incoming magnetization current in the right lead, the incoming magnetization current vector can be written as
\begin{align}
\bm{I}= \frac{\beta V}{\hbar} ( 1, 0, 1, 0, \cdots 1, 0) \equiv \frac{\beta V}{\hbar} \bm{v}\, ,
\end{align}
\noindent and the conductance \(G\) \cite{Meier2003_MagnetizationTransport} is given by 
\begin{align}
G \Delta B = \sum_{l=1}^N (O_l^L - I_{l}^R) = \frac{(g \mu_B)^2 \Delta B }{\pi \hbar}  \bm{v}^T S \bm{v},
\label{eq_conductance_matrixExpression}
\end{align}
\noindent where we have restored the reduced Planck constant \(\hbar\). We perform the explicit calculation of the matrix product \(\bm{v}^T S \bm{v}\) 
in Appendix~\ref{app_conductance}. The result gives
\begin{align}\label{eq:conductance}
G =  \frac{(g \mu_B)^2 }{\pi \hbar} \left(\frac{\sigma}{2n } \right) \frac{ (\sigma + 2n )^N  - (\sigma - 2n)^N}{(\sigma + 2n)^{N} + (\sigma - 2n)^{N} },
\end{align}
\noindent and in the limit \(N \rightarrow \infty\), we obtain the spin conductance
\begin{align}
G =  \frac{(g \mu_B)^2 }{\pi \hbar} \cdot \frac{\sigma}{2n} \equiv \frac{\sigma}{2n}G_0,
\end{align}
\noindent where $G_0$ is the quantum of conductance. %\et{[Check]}
Thus, the spin fractionalization is also manifest in the quantized two-terminal spin conductance \(G\).

\subsection{{Chiral spin liquids}} \label{sec_fsqhe_csl}

The existence of chiral edge states ensures that the phases we find have non-local chiral order. 
We now check that the phases also have a non-trivial local chiral order parameter.

To express the local chiral order parameter \(\chi_l(x)\) in Eq.~\eqref{eq_toyModel_chiralOrderParam} in terms of the bosonic fields, we first write the spin operator components in terms of the operators introduced in Eqs.~\eqref{eq:T_LR} and~\eqref{eq:T_W} through 
\begin{subequations} \label{eq:s_alpha}
\begin{align}
    S^{z}_{j,l}&\sim M_l +  \frac{1}{ {2\pi }} \left[ \left(T_l^L\right)^\dagger T_l^R + \mathrm{H.c.}  \right], \\
    S^{x}_{j,l}&\sim \frac{1}{\sqrt{2\pi } } \left[ T_l^L+\left(T_l^L\right)^\dagger+T_l^R+\left(T_l^R\right)^\dagger \right] T_l^{W,S}, \\
    S^{y}_{j,l}&\sim \frac{1}{\sqrt{2\pi } } \left[ T_l^L-\left(T_l^L\right)^\dagger+T_l^R-\left(T_l^R\right)^\dagger \right] T_l^{W,S},
\end{align}
\end{subequations}
where $M_l=\langle S^{z}_{j,l} \rangle$, again, is the $z$-magnetization of chain $l$. 
All operators \(T_l^\alpha\) are evaluated at $x=ja$, where $a$ is both the cutoff and the lattice spacing and we let $a\to 0$. Finally, we have disregarded terms proportional to $\partial_x \phi_l$ or $\partial_x \theta_l$,
as they later average to zero.  We obtain 
\begin{equation}\label{eq:chi2}
    \begin{split}
        \chi_l(x) 
        = \sum_{\{\tilde{s}, \tilde{\kappa}\}} C_{\{\tilde{s},\tilde{\kappa}\}} \prod_{\delta=0}^1 ( T_{l+\delta}^W )^{\tilde{\kappa}_l^\delta} (T_{l+\delta}^R)^{\tilde{s}_l^{R,\delta} } (T_{l+\delta}^L)^{\tilde{s}_l^{L,\delta} },
    \end{split}
\end{equation} 
with integers \(\tilde{s}_l^{\alpha,\delta}\) such that \(\tilde{s}_l^{\alpha, \delta} \in \{0, \pm 1, \pm 2\}\).
With \(\tilde{\sigma}_l^\delta = \tilde{s}_l^{L,\delta} + \tilde{s}_l^{R,\delta}\), we further require \(|\tilde{\sigma}_l^\delta| \in \{0,1\}\), \(\sum_\delta \tilde{\sigma}_l^\delta = 0\), $\lvert \tilde{s}_l^{L,\delta} \rvert + \lvert \tilde{s}_l^{R,\delta} \rvert \leq 3$, and $\sum_\delta \left(\lvert \tilde{s}_l^{L,\delta} \rvert + \lvert \tilde{s}_l^{R,\delta} \rvert \right)\leq 4$. The integers $\tilde{\kappa}_{l}^\delta$ are given by 
\(\tilde{\kappa}_l^\delta =\pm 1\) when \(\tilde{\sigma}_l^\delta\) is odd and \(\tilde{\kappa}_l^\delta = 0\) when \(\tilde{\sigma}_l^\delta\) is even. 
The prefactors $C_{\{\tilde{s},\tilde{\kappa}\}}$ are in general complex.
The term obtained from Eq.~\eqref{eq:chi2} by letting $(\tilde{s}_l^{L, 0}, \tilde{s}_l^{R, 0}, \tilde{s}_l^{L, 1}, \tilde{s}_l^{R, 1} ) = (0, -1, 1, 0)\) and \((\tilde{\kappa}_l^0, \tilde{\kappa}_l^1) = (1, 1)\) (in addition to the Hermitian conjugate) coincides with the effective interaction in Eq.~\eqref{eq_fqheInteraction} when $(\sigma, n) = (1, 1)$. 
When this interaction dominates the physics, $\langle \chi_l(x) \rangle$ therefore develops a finite expectation value.

For most other values of $\sigma$ and $n$, the effective interaction responsible for the gap is not present in \(\chi_l(x)\) above.
This simply means that the local chiral order parameter characterizing these phases is a generalization of \(\chi_l(x)\). 
Such a generalization may involve the product of a few different $\chi_{ijk}$ from triplets of nearby spins.
Since the gap arises from the chiral interaction only, we do not expect any spontaneous formation of a magnetic pattern in the ground state. In conclusion, we have a phase with no spontaneous magnetic order but instead with local and non-local chiral order, which is therefore a CSL phase.

\section{Higher level FSQHE phases}\label{sec_higherLevel}
The FSQHE phases discussed so far have one chiral edge state and correspond to first-level states in the hierarchy classification of standard quantum Hall states~\cite{Haldane1983_Fractional, Halperin1984_Statistics}. We construct second-level states with two chiral edge states by considering 
an interaction involving three neighbouring spin chains and letting  
\(\sigma_l^1 = - \sigma_l^{-1} = \sigma\), \(\sigma_l^0 = 0\), \(n_l^0 = n_0\), and \(n_l^1 = n_l^{-1}= n_1\), while all other  \(\sigma_l^\delta\) and \(n_l^\delta\) are set to zero~\cite{Teo2014_From}. 
Here, we have introduced integers \(\sigma\), \(n_0\), and \(n_1\) specifying the interaction, as also given in Table~\ref{tab_integerDefinitions}. 

In the following, we first show why the above interaction gives rise to a gapped bulk and two chiral edge states on each edge. We then show how the magnetizations and DMI strengths can be engineered to make the interaction resonant. We provide constructions for the two cases \((\sigma, n_0, n_1) = (1, 1, 0)\) and \((\sigma, n_0, n_1) = (1, 1, 1)\). We then derive the weak coupling phase diagram for the former. Interestingly, this phase can be made renormalization group relevant even at weak coupling for a large parameter space region.
%We then discuss the properties of the phase dominated by interactions of the form above
We then calculate the fractional spin of the basic excitations as well as the conductance. For the phase with \((\sigma, n_0, n_1) = (1, 1, 0)\), we find that the basic spin excitations carry spin \(s^* = 1\). Despite belonging to the second-level FSQHE hierarchy, the phase can therefore not be considered truly fractional. On the other hand, for \((\sigma, n_0, n_1) = (1, 1, 1)\), the phase hosts excitations with fractional spin \(s^* = 1/3\). 

\subsection{Bulk gap}
The bosonic field \(\varphi_l^\mathrm{2nd}\) associated with the interaction is given by 
\begin{align}
\varphi_l^\mathrm{2nd} = \sigma (\theta_{l+1} - \theta_{l-1} ) + 2n_0 \phi_l + 2n_1( \phi_{l-1} + \phi_{l+1}).
\label{eq_higherLevel_bosonicField}
\end{align}
The interaction term corresponding to the second-level FSQHE phase then takes the form of Eq.~\eqref{eq_fqheInteraction} with \(\varphi_l \rightarrow \varphi_l^\mathrm{2nd}\) 
and where the summation index $l$ instead runs from $2$ to $N-1$. 

The interaction describes the net scattering of \(\sigma\) particles from chain \(l-1\) to chain \(l+1\) and backscattering of particles on the chains \(l-1\), \(l\), and \(l+1\) as determined from the integers \(s_{l}^{\alpha, \delta}\) obtained through Eq.~\eqref{eq_sParameters}. For instance, when \((\sigma, n_0, n_1) = (1, 1, 0)\),
two integers \(\sigma_l^\delta\) are odd. Thus, there are four different choices for \(\{\kappa_l^\delta\}\), corresponding to four types of scattering processes leading to the same bosonic interaction, 
as shown in Fig.~\ref{fig_suppMat_scatteringProcesses}(b). 
The lowest order contributions are proportional to \((J_\perp^{xy})^2\). 

From the commutation relations in Eq.~\eqref{eq_commutationBosonFields}, it follows that the fields \(\varphi_l^\mathrm{2nd}(x)\) and \(\varphi^\mathrm{2nd}_{l'}(x')\) commute for any \(l\), \(l'\), and \(x\), \(x'\), and the fields in the set \(\{\varphi_l^\mathrm{2nd}\}\) 
can therefore order independently.
To understand the nature of the phase in the strong coupling regime, we introduce the bosonic fields 
\begin{subequations} \label{eq_higherLevel_chiralFields}
\begin{align}
\eta_{2k-1}^R &=  \sigma \theta_{2k-1} - 2 n_1 \phi_{2k-1} - 2 n_0 \phi_{2k}, \\
\eta_{2k-1}^{L} &=  \sigma \theta_{2k-1} + 2 n_1 \phi_{2k-1}, \\
\eta_{2k}^R \,\,\,\,\, &=  \sigma \theta_{2k}  - 2 n_1 \phi_{2k}, \\
\eta_{2k}^L \,\,\,\,\, &=  \sigma \theta_{2k} + 2 n_1 \phi_{2k} + 2 n_0 \phi_{2k-1},
\end{align}
\end{subequations}
\noindent where \(k\) is an integer taking values \(k = 1, ..., N/2\). Here, \(N\) is the total number of spin chains, which we assume to be an even number. 
Using Eq.~\eqref{eq_higherLevel_chiralFields}, we have $\varphi_l^\mathrm{2nd} =  \eta_{l+1}^L- \eta_{l-1}^R$, such that
\begin{align} \label{eq_higherLevel_interactionTerm}
\mathcal{H}^\mathrm{int}= \sum_{l=2}^{N-1} \int \frac{dx}{2\pi} g_l \cos (\eta_{l+1}^L - \eta_{l-1}^R ).
\end{align}
\noindent When the interaction is relevant, it gaps out $2 (N-2)$ modes, while leaving the chiral modes  \(\eta_{1}^L\), \(\eta_2^L\), \(\eta_{N-1}^R\), and \(\eta_{N}^R\) gapless.
%two chiral modes on the left edge, \(\eta_{1}^L\) and \(\eta_2^L\), and two on the right edge, \(\eta_{N-1}^R\) and \(\eta_{N}^R\), respectively, gapless.

\subsection{Explicit construction of second-level phases}

In this section, we give two examples of how to construct second-level phases; for \((\sigma, n_0, n_1) = (1, 1, 0)\) and for \((\sigma, n_0, n_1) = (1, 1, 1)\). 
We emphasize that there are many different ways to choose the magnetizations and DMI strengths in order to make the bosonic interaction terms resonant, and the specific examples below are meant to illustrate that choosing suitable system parameters such that the target interaction becomes resonant is possible.

\subsubsection{Construction of phase with \texorpdfstring{\((\sigma, n_0, n_1) = (1, 1, 0)\)}{sigma=n0=1 and n1=0}}

To make an explicit suggestion for a construction where the interaction specified by \((\sigma, n_0, n_1) = (1, 1, 0)\) is resonant, we consider a system with a unit cell composed of six chains. The resonance condition then takes the form of Eq.~\eqref {eq_matrixResonanceCondition}, where the matrices \(A\) and \(D\) are now given by 
\begin{align}
A = \begin{pmatrix}
n_0 & n_1 & 0 & 0 & 0 & n_1 \\
n_1 & n_0 & n_1 & 0 & 0 & 0 \\ 
0 & n_1 & n_0 & n_1 & 0 & 0 \\
0 & 0 & n_1 & n_0 & n_1 & 0\\
0 & 0 & 0 & n_1 & n_0 & n_1 \\
n_1 & 0 & 0 & 0 & n_1 & n_0 \\
\end{pmatrix},
\label{eq_resonanceConditionMatrices_A}
\end{align}
\begin{align}
D = \sigma \begin{pmatrix}
0 & -1 & 0 & 0 & 0 & 1 \\
1 & 0 & -1 & 0 & 0 & 0  \\
0 & 1 & 0 & -1 & 0 & 0   \\
0 & 0 & 1 & 0 & -1 & 0   \\
0 & 0 & 0 & 1 & 0 & -1  \\
-1 & 0 & 0 & 0 & 1 & 0 \\
\end{pmatrix}.
\label{eq_resonanceConditionMatrices_D}
\end{align}
For uniform \(p_l = p\) and \((\sigma, n_0, n_1) = (1, 1, 0)\), \( (A^{-1} \bm{p} )_l = p\) so that any choice of non-zero \(\tilde{\bm{b}}\) prevents \(f_l \in (0, 1)\) for at least one $l$. 
To have a valid construction for the simplest second-level phase, we therefore need an inhomogeneous integer valued column vector \(\bm{p}\).

\begin{figure}
    \centering
    \hspace{-0.7cm}
    \includegraphics[width=0.94\columnwidth]{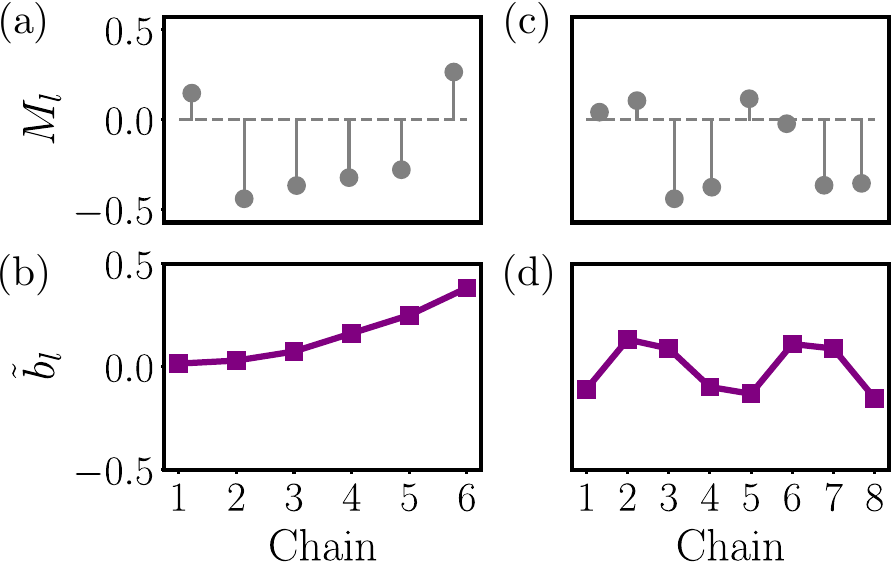}
    \caption{
    Magnetization and dispersion relation shift profiles for the construction of second-level FSQHE phases. When the interaction is relevant, the corresponding phase hosts two chiral gapless modes on each edge. (a-b) Construction of phase with \((\sigma, n_0, n_1) = (1, 1, 0)\). (c-d) Construction of phase with \((\sigma, n_0, n_1) = (1, 1, 1)\).}
    \label{fig_hierarchyProfile}
\end{figure}

Choosing \(\bm{p}\) and \(\tilde{\bm{b}}\) given by  
\begin{align}
{\bm{p}} &= \begin{pmatrix}  1, & 0, & 0, & 0, & 0, & -1 \end{pmatrix}^T,
\\
\tilde{\bm{b}} &= \begin{pmatrix} 1, & 2, & 5, & 11, & 17, & 26 \end{pmatrix}^T / 68,
\end{align}
we obtain physical filling fractions \(f_l\) given by 
\begin{align}
\bm{f} = \begin{pmatrix} 44, &  4, &  9, & 12, & 15, & 52 \end{pmatrix}^T / 68 .
\end{align}
For this choice of filling fractions and dispersion relation shifts, no simpler and more relevant interactions satisfy the resonance condition. We plot the profiles of $M_l$ and $\tilde{b}_l$ for the chosen second-level FSQHE phase construction in Fig.~\ref{fig_hierarchyProfile}(a-b). \\

\subsubsection{Construction of phase with \texorpdfstring{\((\sigma, n_0, n_1) = (1, 1, 1)\)}{n0=n1=1}} 

In this section, we discuss how one may choose the dispersion relation shifts \(\tilde{\bm{b}}\) and filling fractions \(\bm{f}\) to make the interaction specified by \((\sigma, n_0, n_1) = (1, 1, 1)\) resonant. 

We consider a repeated basis of eight chains. The filling fractions and dispersion relation shifts therefore need to satisfy a resonance condition of the form in Eq.~\eqref{eq_matrixResonanceCondition}, where  \(\bm{f}\),  \(\bm{p}\), and \(\tilde{\bm{b}}\) are vectors with eight elements, while the matrices \(A\) and \(D\) are the natural generalizations of the \(6\times 6\) matrices in Eqs.~\eqref{eq_resonanceConditionMatrices_A} and \eqref{eq_resonanceConditionMatrices_D} to matrices of size \(8 \times 8\). The determinant of the \(8\times 8\) matrix \(A\) is
\(\operatorname{det} A = n_0^2 (n_0^2 - 2 n_1^2 )^2 (n_0^2 - 4 n_1^2)\), and the matrix is invertible for \((n_0 , n_1) = (1,1)\). 
Choosing \(p_l = 1\) and 
\begin{align}
%\tilde{\bm{b} } = \begin{pmatrix} 14, &  11, &  13, &   1, &   -8, & -14, &   8, &   5, \end{pmatrix}^T / 95,
%\tilde{\bm{b}} =  \begin{pmatrix} -6 &   1 &  4 &  7 & 5 & -11 &  13 &  14 \end{pmatrix}^T / 97,
\tilde{\bm{b}} = \begin{pmatrix} -10, & 12, & 8, & -9, & -12, & 10, & 8, & -14 \end{pmatrix}^T / 91,
\end{align}
\noindent the resonance condition is satisfied for filling fractions
\begin{align}
%\bm{f} = \begin{pmatrix} 65 & 131 & 86&  38 & 98 & 104 & 131 & 107 \end{pmatrix} / 285.
%\bm{f} = \begin{pmatrix} 16 & 100 & 205 &  4 & 85 & 148 & 82 & 136 \end{pmatrix} / 291.
\bm{f} = \begin{pmatrix} 148, & 166, & 13, & 31, & 169, & 130, & 34, & 37 \end{pmatrix}^T / 273 .
\end{align}
The corresponding magnetization and dispersion relation shift profiles are shown in Fig.~\ref{fig_hierarchyProfile}(c-d). Following the procedure outlined in Sec.~\ref{sec_fsqhe_choice}, we have checked that no simpler and more relevant interactions satisfy the linear and homogeneous Diophantine equations which follow from the resonance condition.

\subsection{Scaling dimension and phase diagram}
\label{sec_higherLevel_scalingDim}

The weak coupling phase diagram of the second-level FSQHE phase can be obtained by calculating the scaling dimension of the interaction %which gives rise to 
responsible for the phase. In the absence of the interchain kinetic coupling in Eq.~\eqref{eq_hPerp_z0}, this scaling dimension is given by 

\begin{align}
\Delta^\mathrm{2nd}_l = \frac{\sigma^2}{4} \left( \frac{1}{K_{l-1}} + \frac{1}{K_{l+1}} \right)
+ {n_0^2} K_l + n_1^2 (K_{l+1} + K_{l-1} ).
\label{eq_scalingDimIntra}
\end{align}

\noindent For simplicity, we again assume chain-independent LL parameters \(K_l = K\). The simplest non-trivial second-level interaction is given by \((\sigma, n_0, n_1) = (1, 1, 0)\). Then, the associated scaling dimension is \(\Delta^\mathrm{2nd} = 1/(2K) + K\), so that the interaction is relevant in the wide regime \(K \in ( 1- 1/\sqrt{2}, 1 + 1/\sqrt{2})\) for the LL parameter. This regime includes significant ferromagnetic and antiferromagnetic intrachain out-of-plane spin interaction~\cite{Giamarchi2003_Quantum}.

More generally, the scaling dimension can also be computed in the presence of the kinetic coupling in Eq.~\eqref{eq_hPerp_z0}, as discussed in Appendix~\ref{app_scalingDimension}.  
We find \(\Delta^\mathrm{2nd} = \Delta^\mathrm{2nd}_\theta + \Delta^\mathrm{2nd}_\phi\) with
\begin{subequations}\label{eq:scal_dim_2}
\begin{align}
\Delta_\theta^\mathrm{2nd}
&= \left( \frac{\sigma^2}{K} \right) [J_0^+(v_\perp^z K) - J_2^+(v_\perp^z K) ], \\
\Delta_\phi^\mathrm{2nd}
&= K [n_0^2 J_0^-(v_\perp^z K) + 4 n_0 n_1 J_1^-(v_\perp^z K) + 4 n_1^2 J_2^-(v_\perp^z K) ].
\end{align}
\end{subequations}
\noindent The result can again be expressed in terms of complete elliptic integrals of the first and second kind. 
Using Eq.~\eqref{eq:scal_dim_2}, we plot the weak coupling phase diagram for the second-level FSQHE phase with \((\sigma, n_0, n_1 ) = (1, 1, 0)\) in Fig.~\ref{fig_phaseDiagram}(b). As expected from the result for \(J_\perp^z = 0\) in  Eq.~\eqref{eq_scalingDimIntra},   the second-level FSQHE phase is relevant in a large portion of parameter space. 
These results show that even very weak interchain interactions may be sufficient to make the simplest non-trivial second-level FSQHE phase relevant. 

Given the existence of chiral edge states in our higher-level FSQHE phases, it should be possible to construct a local chiral order parameter. Based on the interaction giving rise to the second-level phases, this local chiral order parameter should involve spins from three neighbouring chains. The phases therefore satisfy the definition of CSLs.
In the example of the second-level FSQHE phase with \((\sigma, n_0, n_1 ) = (1, 1, 0)\), the associated symmetrized local chiral order parameter $\mathscr{\chi}_{l}^{(1, 1, 0)}(x)$ is
\begin{align}%\label{eq_toyModel_chiralOrderParam}
%\mathscr{\chi}_{l}^{(1, 1, 0)}(x_j) = (\bm{S}_{j,l-1} - \bm{S}_{j+1,l+1} ) \cdot \bm{S}_{j+1,l-1} \times \bm{S}_{j,l+1}.
\mathscr{\chi}_{l}^{(1, 1, 0)}(x_j) = (\bm{S}_{j+1,l} + \bm{S}_{j-1,l} ) \cdot \bm{S}_{j,l-1} \times \bm{S}_{j,l+1}.
\end{align}
%where we simply used Eq.~\eqref{eq_toyModel_chiralOrderParam}, but replaced the label $l$ by the label $l-1$ on the right hand side of the equation, without changing the label $l+1$.

\subsection{Fractional spin}\label{sec_higherLevel_spin}
The spin $s^*$ associated with any excitation is given by Eq.~\eqref{sec_fracSpin_def}.
We now compute $s^*$ for an elementary bulk excitation in a second-level phase. This excitation is given by a single jump from 0 to $\pm 2\pi$ in the bosonic field  $\varphi^{2\textrm{nd}}_{l^*}$ associated with a single index $l^*$. To calculate the spin associated with the basic spin excitation, we first relate \(\sum_l \phi_l\) to the fields \(\eta_{l}\) in Eq.~\eqref{eq_higherLevel_chiralFields}. Assuming $4n_1^2 \neq n_0^2$ to have a non-singular relation between the chiral fields \(\eta^{L}, \eta^{R}\) and \(\theta, \phi\), we find
\begin{equation}
\begin{pmatrix} \theta_{2k-1} \\ \phi_{2k-1} \\ \theta_{2k} \\ \phi_{2k} \end{pmatrix}
= \frac{1}{\mathcal{N}}
\begin{pmatrix}
\frac{4n_1^2-2n_0^2}{\sigma} & \frac{4n_1^2}{\sigma} & \frac{2n_1 n_0 }{\sigma} & -\frac{2n_1 n_0 }{\sigma}  \\
2 n_1& -2n_1 & -n_0 & n_0 \\
-\frac{2n_1 n_0 }{\sigma} & \frac{2n_1 n_0 }{\sigma}  & \frac{4n_1^2}{\sigma} & \frac{4n_1^2-2n_0^2}{\sigma}\\
-n_0 &  n_0 & 2n_1 & -2n_1 
\end{pmatrix}
\begin{pmatrix} \eta_{2k-1}^L \\ \eta_{2k-1}^{R} \\ \eta_{2k}^L \\\eta_{2k}^R \end{pmatrix}
\end{equation}
\noindent with prefactor \(\mathcal{N} = 2(4n_1^2-n_0^2)\) and \(k =1,..., N/2\). Consequently, we find
\begin{align}
s^* &=
- \frac{1}{n_0 + 2n_1} \sum_{l=1}^{N} \int \frac{dx}{2\pi} \; \partial_x \left( \eta_{l}^L-\eta_{l}^R \right).
\end{align}
Rewriting $s^*$ in terms of the fields \(\varphi_l^\mathrm{2nd}\), we then get 
\begin{align}
s^* =- \frac{1}{n_0 + 2n_1} 
%\int \frac{dx}{2\pi} \;  \left( \partial_x \varphi_\mathrm{edge} + \sum_{l=2}^{N-1} \partial_x   \varphi^{2\textrm{nd}}_l \right),
\sum_{l=2}^{N-1} \int \frac{dx}{2\pi} \;   \partial_x   \varphi^{2\textrm{nd}}_l ,
\end{align}
where we have disregarded the edge contribution from the field derivative \( \partial_x(\eta_{N}^R+\eta_{N-1}^R-\eta_{1}^L-\eta_{2}^L)\).
Since we consider a kink in a given bulk bosonic field \(\varphi_{l^*}^\mathrm{2nd}\), we get  spin \(s^* = 1/(n_0 + 2 n_1)\), and the elementary spin excitations are, in general, fractional.

For the case \((\sigma, n_0, n_1) = (1, 1, 0)\), we have spin \(s^* = 1\). For the case \((\sigma, n_0, n_1) = (1,1,1)\), we have fractional spin \(s^* = 1/3\). Thus, the second-level phases can have excitations with both integer and fractional spin.

\subsection{Spin conductance}\label{sec_higherLevel_spinConductance}

\begin{figure}
    \centering
    \includegraphics[width=0.85\columnwidth]{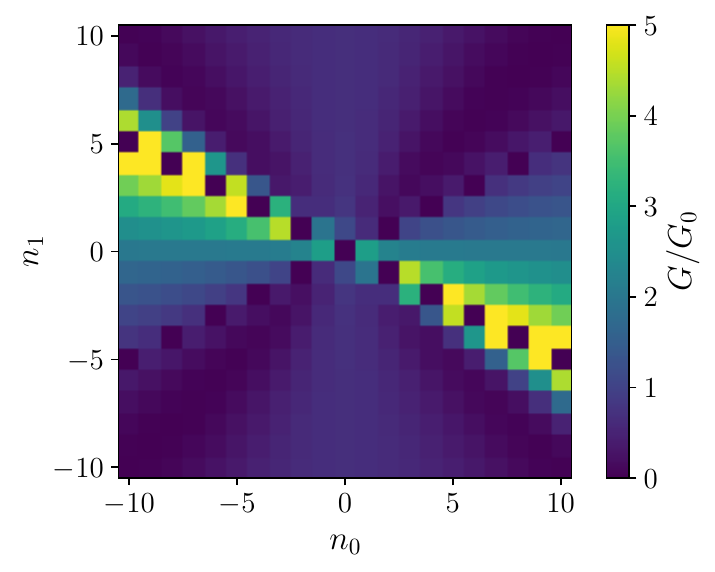}
    \caption{Spin conductance for the second-level FSQHE phase in the limit \(N \rightarrow \infty\) in units of \(G_0 = (g \mu_B)^2 /\pi \hbar\). The conductances, calculated numerically, are fractional numbers in units of $G_0$. }
    \label{fig_conductance2nd}
\end{figure}

We compute the spin conductance in the second-level phase
by the same procedure as in Sec.~\ref{sec_fsqhe_conductance}.
The conductance is
given by 
Eq.~\eqref{eq_conductance_matrixExpression}, where the matrix \(S\) is given by \(S = A^{-1}B\), but the definition of \(A\) and \(B\) is different since the relation between the fields \((\eta_l^{L}, \eta_l^R)\) and \((\phi_l,\ \theta_l)\) has changed. 
We then calculate the conductance numerically. The result is given in Fig.~\ref{fig_conductance2nd}, where the phases with \(n_0^2 = 4 n_1^2\) have been excluded, as they render the relation between the incoming and outgoing currents singular and prevent the calculation of the conductance. 
Clearly, the conductance depends crucially on the integers \(n_0\) and \(n_1\) specifying the various second-level FSQHE phases hosting excitations with different spin \(s^*\).

One may also derive a cumbersome analytical expression for the conductance.
For the special case of \(n_1 = 0\) in the limit of \(N \rightarrow \infty\), it simplifies to
\begin{align}
G|_{n_1 = 0} =\frac{(g \mu_B)^2 }{\pi \hbar} \frac{2}{n_0}  \sqrt{n_0^2+ \sigma^2 }.
\end{align}
While the spin conductance depends strongly on the integers \(n_0\) and \(\sigma\) specifying the interaction, the relation between conductance and the fractional spin is not as straightforward as for the first-level phase.

\section{Breaking the conservation of magnetization}\label{sec_breakingMagnetizationConservation}

In our model, we have assumed \(J_l^x = J_l^y\), \(J_\perp^x = J_\perp^y\), and that the magnetic field and DMI vector are oriented along the direction \(\hat{z}\). Together, this ensures the conservation of the total magnetization operator \(S^z = \sum_{l,j} S_{j,l}^z\). After the Jordan-Wigner transformation, the magnetization conservation constraint corresponds to the conservation of the particles created and annihilated by \(c_i^\dagger\) and \(c_i\). After the bosonization transformation, this gives rise to the constraint \(\sum_\delta \sigma_l^\delta = 0\). 

In this paper, we suggest engineering the wire magnetizations and DMI strengths so that a particular target interaction becomes resonant and more relevant than any other interaction. Breaking the conservation of magnetization therefore allows more competing interactions, and in turn makes it more difficult for the target interaction to dominate the physics. One should therefore consider whether the suggested phases are stable against perturbations breaking the conservation of the magnetization along $z$-direction. 

For concreteness, we assume that conservation of magnetization is broken by a transverse magnetic field term \(H_x = h_x \sum_{l,j} S_{j,l}^x\) and treat this magnetic field term as a perturbation. The perturbatively generated interactions can then still be written in the form of Eq.~\eqref{eq:inter_1stform}, but where the interaction now does not need to satisfy the constraint \(\sum_\delta \sigma_l^\delta = 0\). Any perturbatively generated interaction that breaks this constraint must necessarily involve at least one term from \(H_x\). Assuming that our target interaction is (renormalization group) relevant and that the symmetry-breaking term is sufficiently weak, the target interaction still dominates the physics because of the larger bare coupling strength. Moreover, the new interaction terms that can be generated in the presence of the transverse magnetic field also need to satisfy the resonance condition. We have already chosen the magnetizations and dispersion relation shifts in the explicit constructions for the various phases such that no interaction terms are more relevant than the target interaction, irrespective of whether they break the magnetization conservation constraint or not. Consequently, the effective low-energy description is left unchanged, and conservation of the $z$-magnetization is restored in the low-energy description of the system.

\section{Discussion}\label{sec_discussion}

To realize our proposed coupled spin chain system hosting the FSQHE phases discussed above, one needs independent control over the interactions characterizing the chains and, in particular, modulated DMI and chain-dependent magnetizations. We suggest that this can be realized in synthetic spin systems such as synthetic spin chains, ultracold atom systems, or Rydberg atoms. We discuss these cases briefly in the following.

Using scanning-tunneling microscopy techniques, it is possible to construct artificial atomic spin chains on various substrates. Incidentally, positioning magnetic atoms on superconductors has been a popular path to search for Majorana bound states~\cite{Heinrich2006, Nadj-Perge2014, Klinovaja2013, Pawlak2016, Kamlapure2018_EngineeringSpinCouplings, Beck2023_SearchLargeTopological, Kim2018_TailoringMajoranaBound, Pawlak2019, Schneider2020_ControllingIngapEnd, Ding2021, Kuster2022, Schneider2023_ProbingTopologicallyTrivial}. 
Using these techniques, a certain control over the spin interactions in the system has been demonstrated~\cite{Heinrich2006,Kamlapure2018_EngineeringSpinCouplings,Ding2021}. A key ingredient for the realization of the FSQHE phase is modulated DMI. In a system of multiple artificial spin chains, this could be engineered by patterning strips of heavy metals with strong spin-orbit interaction on the substrate~\cite{Hill2017_Spin}. Proposals also exist to control and modulate the DMI strength through electric fields~\cite{Yang2018_Controlling} or laser irradiation~\cite{Stepanov2017_Dynamical, Losada2019_Ultrafast}. Position-dependent magnetizations can be obtained through inhomogeneous magnetic fields produced e.g. by magnetic domains~\cite{Desjardins2019_Synthetic}, 
but tuning the magnetization of individual chains with the precision required for the FSQHE remains a challenge. 

In multi-component ultracold atom systems, one may exploit Feshbach resonances to enter a Mott insulating regime~\cite{Greiner2002_Quantum, Chin2010_Feshbach}. The low-energy excitations can then be described by an effective spin model.  Utilizing laser coupling, one may achieve control over the spin interactions and realize modulated DMI~\cite{Galitski2008_SpinOrbit, Lin2011_SpinOrbit, Su2015_Position} and effective Zeeman fields~\cite{Lim2010_Artificial}. Furthermore, by changing the ratio of the lateral to longitudinal lattice constants, it is possible to enter a regime where the spin chains are weakly coupled. Ultracold atoms may therefore also provide a suitable platform to explore the FSQHE in weakly coupled spin chains. In addition to the proposal of this paper, multicomponent ultracold atoms have also been proposed to realize effective quantum spin Hamiltonians hosting chiral spin liquid phases through spontaneous symmetry breaking~\cite{Sedrakyan2015_Spontaneous, Hermele2009_MottInsulators} and the chiral spin interaction~\cite{Nielsen2013, Hickey2015_Competing, Hickey2016_HaldaneHubbard}. 

Rydberg atoms can be trapped to form Rydberg atom lattices, which can again simulate a number of condensed matter phenomena~\cite{Browaeys2020_ManyBodyPhysicsWithRydberg}. These systems have also been proposed as candidates to host chiral spin liquids~\cite{Tarabunga2023_Classification}. Motivated by the realization of spin-orbit coupling in ultracold atom systems, it has furthermore been proposed that a controllable DMI interaction can be realized in Rydberg atom systems using two-photon Raman transitions~\cite{Kunimi2023_Proposal}, which is a key ingredient also in our proposal.

\section{Summary}\label{sec_summary}
Starting from coupled spin chains subject to chain-dependent magnetic fields and Dzyaloshinskii-Moriya coupling strengths, we showed that well-engineered interactions involving the transfer of spin between neighbouring chains give rise to CSL phases, which are spin analogues of the fractional quantum Hall effect in two-dimensional electron gases. This is achieved through modulated intrachain DMI
and independent control of the chain magnetizations. We demonstrated this through the explicit construction of first-level and second-level phases within the fractional quantum Hall hierarchy. Within these phases, the elementary spin excitations carry fractionalized spin, and the spin fractionalization is
manifest in the spin conductance. 
We propose that realizing these phases could be achievable in systems of synthetic spin chains or ultracold atoms.

\section{Acknowledgements}
 This project has received funding from the European Union's Horizon 2020 Research and Innovation Program under Grant Agreement No. 862046 and under Grant Agreement No. 757725 (the ERC Starting Grant). This work was supported by the Swiss National Science Foundation.
 %, and NCCR SPIN, a National Centre of Competence (or Excellence) in Research, funded by the Swiss National Science Foundation (Grant No. 51NF40-180604).

\appendix

\section{{Bosonization of the local chiral order parameter}} \label{app_bosoChi}

To bosonize $\chi_{nmk}$, where $n=(j,l)$ (site $x_j=ja$ on chain $l$), $m=(j+1,l)$, and $k=(j,l+1)$, we start from Eq.~\eqref{eq:s_alpha}. We then replace the operators $T^W_l$, $T^R_l$, and $T^L_l$ with their expression in terms of bosonic fields given by Eqs.~\eqref{eq:T_LR} and ~\eqref{eq:T_W}.
We introduce $\mathcal{L}_l$ and $\mathcal{R}_l$ such that $T_l^L=\nu_L e^{i\mathcal{L}_l}$, $T_l^R=\nu_R e^{i\mathcal{R}_l}$, and $T_l^W=e^{\frac{i}{2}(\mathcal{L}_l-\mathcal{R}_l)}$ to shorten the notation, so that $\mathcal{R}_l(x)=k_l^R x + \theta_l(x) - \phi_l(x)$ and $\mathcal{L}_l(x)=k_l^L x + \theta_l(x) + \phi_l(x)$. We keep the Klein factors $\nu_L$ and $\nu_R$ to find the correct signs. The derivatives of the fields $\mathcal{L}$ and \(\mathcal{R}\) are now related to the fields \(\theta\) and \(\phi\) through 
\begin{align}
&\partial_x \mathcal{L}_{l} - \partial_x \mathcal{R}_{l} = -2 k_l + 2 \partial_x \phi_l, \\
%= - \frac{2\pi}{{a}} \left(M_l + \frac{1}{2} \right) + 2 \partial_x \phi_l, \\
&\partial_x \mathcal{L}_{l} + \partial_x \mathcal{R}_{l} = \,\,\,\, 2 b_l + 2 \partial_x \theta_l,
\end{align}
where, as in the main text, $k_l$ is the Fermi momentum and $b_l$ is the momentum shift. We can then express the spins in terms of \(\mathcal{L}\) and \(\mathcal{R}\) through 
\begin{align}
&S^x_{j,l} = \frac{1}{\sqrt{2\pi}}\left[\nu_L\cos(\mathcal{L}_{l})+ \nu_R\cos(\mathcal{R}_{l})\right]\cos[(\mathcal{L}_l-\mathcal{R}_l)/2],\label{eq:Sx_boso}\\
&S^y_{j,l} = -\frac{1}{\sqrt{2\pi}}\left[\nu_L\sin(\mathcal{L}_{l})+ \nu_R\sin(\mathcal{R}_{l})\right]\cos[(\mathcal{L}_l-\mathcal{R}_l)/2], \label{eq:Sy_boso} \\
&S^z_{j,l} = {M_{l}} -\frac{a}{\pi}\partial_x \phi_l- i\nu_R\nu_L  \left(\frac{k_l a}{\pi}-\frac{a}{\pi}\partial_x \phi_l \right) \cos (\mathcal{L}_{l}-\mathcal{R}_{l}), \label{eq:Sz_boso}
\end{align}
where $M_l$ is the magnetization. We now write 
\begin{equation}\label{eq_app_chiralUnsymm}
\bm{S}_{j,l} \cdot \bm{S}_{j+1,l} \times \bm{S}_{j,l+1} = \sum_{n} X_{j,l}^n,
\end{equation}
\noindent where the different terms \(X_{j,l}^n\) contributing to the unsymmetrized chiral order parameter above are given by
\begin{align}
X_{j,l}^1 = S^z_{j,l+1}[S^x_{j,l}S^y_{j+1,l}-S^y_{j,l}S^x_{j+1,l}], \\
X_{j,l}^2 = S^x_{j,l+1}[S^y_{j,l}S^z_{j+1,l} -S^z_{j,l}S^y_{j+1,l}], \\
X_{j,l}^3 = S^y_{j,l+1}[S^z_{j,l}S^x_{j+1,l}-S^x_{j,l}S^z_{j+1,l}].
\end{align}
Introducing the bosonized representation for the spin operators, we obtain
\begin{subequations} \label{eq:chi_boso1}
\begin{align}
X_{j,l}^1 =\sum_{\alpha_{l+1},\beta_l}C_{\alpha_{l+1},\beta_l}\cos(\alpha_{l+1}+\beta_l + \Phi_C), \\
X_{j,l}^2 + X_{j,l}^3 =\sum_{\gamma_{l+1},\delta_l}D_{\gamma_{l+1},\delta_l}\cos(\gamma_{l+1}-\delta_l + \Phi_D), \label{eq:chi_term_cross}
\end{align}
\end{subequations}
where $\alpha_{l+1} \in \{ 0, \mathcal{L}_{l+1}-\mathcal{R}_{l+1}\}$ and $\beta_l=\pm(\mathcal{L}_{l}-\mathcal{R}_{l})$, while for the second pair of fields, we have $\gamma_{l+1} \in \{(\mathcal{L}_{l+1}+\mathcal{R}_{l+1})/2, (3\mathcal{L}_{l+1}-\mathcal{R}_{l+1})/2$, $(3\mathcal{R}_{l+1}-\mathcal{L}_{l+1})/2 \}$ and $\delta_l \in\{(\mathcal{L}_{l}+\mathcal{R}_{l})/2, (3\mathcal{L}_{l}-\mathcal{R}_{l})/2$, $(3\mathcal{R}_{l}-\mathcal{L}_{l})/2, (5\mathcal{L}_{l}-3\mathcal{R}_{l})/2, (5\mathcal{R}_{l}-3\mathcal{L}_{l})/2\} $. The coefficients $C_{\alpha_{l+1},\beta_l}$ and $D_{\gamma_{l+1},\delta_l}$ consist of real numbers and field derivatives, and \(\Phi_C\) and \(\Phi_D\) are some unimportant phase shifts. 

Similarly, we may now derive the bosonized form of the unsymmetrized chiral interaction for any other triplet of spins from the smallest plaquette on the square lattice, and subsequently exploit this to derive the expression for any symmetrized chiral order parameter. In particular, we may consider the symmetrized chiral order parameter \(\chi_l(x)\) in Eq.~\eqref{eq_toyModel_chiralOrderParam}. The result is of the form 
\begin{align}
\chi_l(x) = & \,\,\,\,\, \sum_{\alpha_{l+1},\beta_l}C_{\alpha_{l+1},\beta_l}\cos(\alpha_{l+1}+\beta_l + \Phi_C) \nonumber \\
&+ \sum_{\alpha_{l}, \beta_{l+1}} {C}_{\alpha_{l},\beta_{l+1}}\cos(\alpha_{l}+\beta_{l+1} + \Phi_C) \nonumber\\
&+ \sum_{\gamma_{l+1},\delta_l}D_{\gamma_{l+1},\delta_l}\cos(\gamma_{l+1}-\delta_l + \Phi_D) \nonumber \\
&+ \sum_{\gamma_{l},\delta_{l+1}}D_{\gamma_{l},\delta_{l+1}}\cos(\gamma_{l}-\delta_{l+1} + \Phi_D) .
\end{align}
While in the main paper, we have picked a given symmetrization for the local chiral order parameter, there are in general numerous choices. Other and more involved ways to symmetrize may have the advantage that some of the terms in the above inventory cancel out.

\section{Scaling dimension}\label{app_scalingDimension}

In this Appendix, we show how to calculate the scaling dimension of an interaction term when the kinetic Hamiltonian takes the form of Eq.~\eqref{eq_fsqhe_scalingDim_Hquad}. We then specialize the general calculation to the special cases of first- and second-level FSQHE interactions.

We consider a field $\mathcal{I}$ specified by the real coefficients \(a_n\) and \( b_n\)
such that \(\mathcal{I} = \exp \left[  i \sum_n (a_n \theta_n + b_n \phi_n) \right]\). Since the cumulant expansion is truncated at second order when the action is Gaussian~\cite{vanKampen1992_Book}, the scaling dimension can be extracted from the correlation functions \(\langle \Delta \phi_n(x) \Delta \phi_m(0) \rangle \) and \(\langle \Delta \theta_n(x) \Delta \theta_m(x) \rangle \), where we have introduced \(\Delta \phi_n(x) = \phi_n(x) - \phi_n(0)\) and \(\Delta \theta_n(x) = \theta_n(x) - \theta_n(0)\). The cross correlators \(\langle \theta_n(x) \phi_m(0) \rangle\) and \(\langle \phi_n(x) \theta_m(0) \rangle\) only contribute to the phase of the correlation function, and there is no need to consider them in the following.

We calculate the scaling dimension of the field \(\mathcal{I}(x)\). As discussed in the main text, the Hamiltonian \(H^\mathrm{quad}_0\) in Eq.~\eqref{eq_fsqhe_scalingDim_Hquad} can be diagonalized by introducing the transformation 
\begin{subequations}
\begin{align}
\theta_n &= \sum_m \Gamma_n U_{nm} \tilde{\theta}_m ,\\
\phi_n &= \sum_m \Gamma_n^{-1} U_{nm} \tilde{\phi}_m ,
\end{align}
\end{subequations}
\noindent where \(\Gamma_n = (u_n K_n /u)^{-1/2}\) and \(U\) is the orthogonal matrix that diagonalizes the matrix \(W_{nm} = (u_n/u)^2 \delta_{nm} + \Gamma_{n}^{-1} v_{nm} \Gamma_m^{-1} \), so that \((U^T W U)_{nm} = d_n \delta_{nm}  \). Introducing the renormalized velocity parameter \(\tilde{u}_m = u \sqrt{d_m}\) and LL parameter \(\kappa_m = 1/\sqrt{d_m}\), the Hamiltonian takes the familiar form
\begin{align}
H = \sum_{m} \int \frac{dx}{2\pi} \tilde{u}_m \left[  
 \kappa_m (\partial_x \tilde{\theta}_{m})^2 
+
\frac{1}{\kappa_m}
(\partial_x \tilde{\phi}_{m})^2 \right] .
\end{align}
\noindent This immediately allows us to write down the correlation functions for the fields \(\tilde{\theta}_n\) and \(\tilde{\phi}_n\), which are the usual LL correlation functions~\cite{Giamarchi2003_Quantum} 
\begin{subequations}
\begin{align}
\langle \Delta \tilde{\theta}_n(\bm{r} ) \Delta \tilde{\theta}_m (\bm{r} ) \rangle &= \frac{1}{2} \delta_{nm} \frac{1}{\kappa_n} \log \left( \frac{r_n^2 + \alpha^2}{\alpha^2} \right) , \\
\langle \Delta \tilde{\phi}_n(\bm{r})  \Delta \tilde{\phi}_m (\bm{r} ) \rangle &= \frac{1}{2} \delta_{nm} \kappa_n \log \left( \frac{r_n^2 + \alpha^2}{\alpha^2} \right) ,
\end{align}
\end{subequations}
\noindent where \(\alpha\) is the short distance cutoff and \(r_n = \sqrt{x^2 +  (\tilde{u}_n \tau)^2}\), where \(\tau\) is imaginary time. This can be exploited to calculate the correlation functions for the original fields, and we obtain
\begin{subequations}
\begin{align}
\langle \Delta \theta_n (\bm{r} ) \Delta \theta_{n'} (\bm{r} ) \rangle &=  
(\Gamma U g^{-} U^T \Gamma)_{nn'} , \\
\langle \Delta \phi_n (\bm{r} ) \Delta \phi_{n'} (\bm{r} ) \rangle &= 
(\Gamma^{-1} U g^+ U^T \Gamma^{-1})_{nn'},
\end{align}
\end{subequations}
\noindent where \(g^{\pm}\) is the diagonal matrix with matrix elements
\begin{align}
g_m^{\pm} = \frac{\kappa_m^{\pm 1}}{2} \log \left( \frac{r_m^2 + \alpha^2}{\alpha^2} \right) 
\end{align}
\noindent on the diagonal. 

From the cumulant expansion, we then obtain that the spatial decay of the correlation function is given by 
\begin{align}
\begin{split}
\langle \mathcal{I}^\dagger(x) \mathcal{I}(0) \rangle &  \propto 
\operatorname{exp} \Big\{
-\frac{1}{2} \sum_{nn'} \Big[ a_n a_{n'} \langle \Delta \theta_{n}(x) \Delta \theta_{n'}(x) \rangle \\
&+ b_n b_{n'} \langle \Delta \phi_n(x) \Delta \phi_{n'}(x) \rangle  \Big] 
\Big\}.
\end{split}
\end{align}
\noindent We then obtain 
\begin{align}
\langle \mathcal{I}^\dagger (x) \mathcal{I}(0) \rangle \propto \frac{1}{x^{2\Delta}},
\end{align}
\noindent with scaling dimension \(\Delta = \Delta_\phi + \Delta_\theta\). Here, the contributions from the correlation functions for \(\theta\) and \(\phi\) are given by 
\begin{subequations}
\begin{align}
\Delta_\theta &= \frac{1}{4} a^T \Gamma U \kappa^{-1} U^T \Gamma  a, \\
\Delta_\phi &= \frac{1}{4} b^T \Gamma^{-1} U \kappa U^T \Gamma^{-1} b ,
\end{align}
\label{eq_app_scalingDim_scalingDimGeneral}%
\end{subequations}
where \(\kappa\) is the diagonal matrix with matrix elements \(\kappa_m\) on the diagonal, while \(a\) and \(b\) are the column vectors composed of the coefficients \(a_n\) and \(b_n\) specifying the field \(\mathcal{I}\). 

We now specialize to the case of periodic boundary conditions, translational invariance (\(K_l = K\) and \(u_l = u\) for all $l$), and nearest neighbour chain coupling, so that \(v_{lm} = (v_\perp^z/2) (\delta_{l,m+1}+ \delta_{l,m-1})\). The matrix \(W_{nm} = (u_n/u)^2 \delta_{nm} + \Gamma_{n}^{-1} v_{nm} \Gamma_m^{-1} \) then reduces to a standard nearest-neighbour hopping Hamiltonian and can be diagonalized by introducing a Fourier transformation. Thus, we obtain eigenvalues \(d_m = 1 + v_\perp^z K \cos p_m\), where $p_m = 2 \pi m/N$ and \(m = 0, 1, 2, \ldots, N-1\). 
Furthermore, the matrix \(U\) diagonalizing \(W\) is given by \(U_{nm}= ({1}/{\sqrt{N}} ) e^{i p_m n}\) and
\begin{align}
(U \kappa^{\pm 1} U^\dagger)_{nn'} = \sum_{m} e^{-i p_m (n'-n) }  \kappa_{m}^{\pm 1}.
\label{eq_app_scalingDim_matrixElement}
\end{align}
In the following, we use this expression to calculate the scaling dimension for the first- and second-level FSQHE interactions. 

\subsection{First-level FSQHE interaction}

For the first-level FSQHE interaction, we have \(a_{n^*+1} = - a_{n^*} = \sigma \) and \(b_{n*} = b_{n^*} = 2n\) for some particular field index \(n^*\), with \(a_n = b_n = 0\) for \(n \neq n^*, n^* +1\). Inserting Eq.~\eqref{eq_app_scalingDim_matrixElement} and \(\Gamma_n = K^{-1/2}\) into Eq.~\eqref{eq_app_scalingDim_scalingDimGeneral}, we then obtain 

\begin{subequations}
\begin{align}
\Delta_\theta^\mathrm{1st} &=  \,\,\,\frac{\sigma^2 }{4K}  \frac{1}{N}\sum_m d_m^{1/2}
\begin{pmatrix} 1 & -1 \end{pmatrix} \begin{pmatrix} 1 & e^{-ip_m} \\ e^{ip_m} & 1 \end{pmatrix} \begin{pmatrix} 1 \\ -1 \end{pmatrix} , \\
\Delta_\phi^\mathrm{1st} &= n^2 K  \frac{1}{N} \sum_m \,\,\,\, d_m^{-1/2}
\begin{pmatrix} 1 & \,\, 1 \end{pmatrix} \begin{pmatrix} 1 & e^{-ip_m} \\ e^{ip_m} & 1 \end{pmatrix} \begin{pmatrix} 1 \\ 1 \end{pmatrix} .
\end{align}
\end{subequations}
\noindent In the limit \(N \rightarrow \infty\), the scaling dimension can be expressed in terms of integrals of the form
\begin{align}
J_n^\pm ( a) &= \int_0^{2\pi} \frac{dp}{2\pi} \cos^n( p) \, (1 + a\cos p)^{\pm 1/2} ,
\end{align}
\noindent and we obtain 
\begin{subequations}
\begin{align}
\Delta_\theta^\mathrm{1st} &= \,\,\,\, \frac{\sigma^2}{2K} [ J_0^+(v_\perp^z K) - J_1^+(v_\perp^z K)] ,\\
%\equiv \left( \frac{\sigma^2}{2K} \right) f_\theta(vK) , \\
\Delta_\phi^\mathrm{1st} &= 2n^2 K [J_0^-(v_\perp^z K) + J_1^-(v_\perp^z K)] .
%\equiv 2 n^2 K f_\phi(vK) .
\end{align}
\end{subequations}

\noindent The integrals \(J_n^\pm\) can be related to the complete elliptic integrals of the first and second kind, i.e. \(K(a)\) and \(E(a)\)~\cite{AbramowitzStegun1965}. In the expressions for the scaling dimensions, we have the particular combinations 
\begin{align}
f_\theta(w) = J_0^+ (w) - J_1^+(w), \\
f_\phi(w) =  J_0^-(w) + J_1^-(w),
\end{align}
and the functions \(f_\phi\) and \(f_\theta\) can be written
\begin{subequations}
\begin{align}
f_\theta(w) &= - \frac{2  \sqrt{1+w}}{3\pi w}  \left[  (1-3w) E\left( \tilde{w} \right)- (1-w) K\left( \tilde{w} \right) \right] , \\
f_\phi(w)  &= \frac{1}{2\pi} \frac{4}{w \sqrt{1+w}} \left[(1 +w)E\left(\tilde{w} \right) - (1-w)K \left( \tilde{w} \right) \right],
\end{align}
\end{subequations}
\noindent where the argument \(\tilde{w}\) of the complete elliptic functions is given by \(\tilde{w}\equiv 2 w/(1+w)\). These expressions are used to compute the phase diagram in Fig.~\ref{fig_phaseDiagram} in the main text.

\subsection{Second-level FSQHE interaction}

For the second-level FSQHE interaction, we similarly obtain 
\begin{subequations}
\begin{align}
\Delta_\theta^\mathrm{2nd} &= \frac{1}{4N K}  \sum_m d_m^{+1/2} 
a^T  \mathcal{M}_3 a,
%\begin{pmatrix} 1 \\ 0 \\ -1 \end{pmatrix}
\\
\Delta_\phi^\mathrm{2nd} &=  \quad \,\, \frac{K}{N} \sum_m d_m^{-1/2} 
b^T \mathcal{M}_3 b,
\end{align}
\end{subequations}
where  \(a = \begin{pmatrix} \sigma & 0 & -\sigma \end{pmatrix}^T\), \(b = \begin{pmatrix} n_1 & n_0 & n_1 \end{pmatrix}^T\), and 
\begin{align}
\mathcal{M}_3 = \begin{pmatrix} 1 & e^{-ip_m} & e^{-2ip_m} \\ e^{ip_m} & 1 & e^{-ip_m} \\ e^{2ip_m} & e^{ip_m} & 1 \end{pmatrix}.
\end{align}
In terms of the functions \(J_n^\pm\), we then obtain 
\begin{align}
\Delta_\theta^\mathrm{2nd} &= \left( \frac{\sigma^2}{K} \right) [J_0^+(v_\perp^z K) - J_2^+(v_\perp^z K) ] \, ,\\
\Delta_\phi^\mathrm{2nd} &= K [n_0^2 J_0^-(v_\perp^z K) + 4 n_0 n_1 J_1^-(v_\perp^z K) + 4 n_1^2 J_2^-(v_\perp^z K) ] .
\end{align}

\section{Calculation of spin conductance}\label{app_conductance}

We now show how the matrix product \(\bm{v}^T A^{-1} B \bm{v}\) in Eq.~\eqref{eq_conductance_matrixExpression} can be computed. The relations between \(O_l^\alpha\) and \(I_l^\alpha\) can be rewritten in the form \(A\bm{O} = B \bm{I}\), where we introduced the vectors \(\bm{O}\) and \(\bm{I}\) with \(2N\) elements through
\begin{align}
\bm{O} &= \begin{pmatrix} O_1^L, & O_1^R, & \cdots & O_N^L, & O_N^R \end{pmatrix}^T, \\
\bm{I} &= \begin{pmatrix} I_1^L, & I_1^R, & \cdots & I_N^L, & I_N^R \end{pmatrix}^T,
\end{align}
\noindent as well as matrices \(A\) and \(B\) of size \(2N \times 2N\). For \(N=3\), they are given by 
\begin{align}
A = \begin{pmatrix}
\gamma_1 & \gamma_2 & 0 & 0 & 0 & 0 \\
\gamma_2 & 0 & \gamma_1 & 0 & 0 & 0 \\
0 & \gamma_1 & 0 & \gamma_2 & 0 & 0 \\
0 & 0 & \gamma_2 & 0 & \gamma_1 & 0 \\
0 & 0 & 0 & \gamma_1 & 0 & \gamma_2 \\
0 & 0 & 0 & 0 & \gamma_2 & \gamma_1 
\end{pmatrix},
\end{align}
\begin{align}
B = \begin{pmatrix}
\gamma_1 & \gamma_2 & 0 & 0 & 0 & 0 \\
0 & \gamma_1 & 0 & \gamma_2 & 0 & 0 \\
\gamma_2 & 0 & \gamma_1 & 0 & 0 & 0 \\
0 & 0 & 0 & \gamma_1 & 0 & \gamma_2 \\
0 & 0 & \gamma_2 & 0 & \gamma_1 & 0 \\
0 & 0 & 0 & 0 & \gamma_2 & \gamma_1 
\end{pmatrix},
\end{align}
\noindent and they are easily generalized to arbitrary \(N\). Here, we introduced \(\gamma_1 = n+\sigma/2\) and \(\gamma_2 = n - \sigma/2\). Direct computation then gives 
\begin{align}
B \bm{v} = (\gamma_1 + \gamma_2) \bm{v} - \gamma_2 (\bm{u}_1 - \bm{u}_{2N}),
\end{align}
\noindent where \(\bm{u}_i\) is the unit vector with \(1\) as its \(i\)-th element, and 0 otherwise. To compute \(\bm{v}^T A^{-1} B \bm{v}\), we therefore need to compute \(\bm{x}^T = \bm{v}^T A^{-1}\), which amounts to finding a solution of \(A^T \bm{x} = \bm{v}\). Within the bulk, the elements of the column vector \(\bm{x}\) with \(2N\) elements are then related through 
\begin{subequations}
\begin{align}
\gamma_2 x_{2k-1} + \gamma_1 x_{2k+1} &= 0, \\
\gamma_1 x_{2k}+ \gamma_2 x_{2k+2} &= 1.
\end{align}
\end{subequations}
\noindent Solving these %difference 
equations gives
\begin{subequations}
\begin{align}
&x_{2k-1} = x_1 (- \gamma_2/\gamma_1)^{k-1}, \\
&x_{2k} = \frac{1}{\gamma_1 + \gamma_2} + \left(x_{2N} - \frac{1}{\gamma_1 + \gamma_2} \right)  \left( - \frac{\gamma_2}{\gamma_1} \right)^{N-k},
\end{align}
\label{eq_app_diffEqSol}%
\end{subequations}
\noindent where \(k = 1, \ldots, N\). The unknown elements \(x_1\) and \(x_{2N}\) are determined through the boundary conditions
%stationary continuity equations at the edges, where we have 
\begin{subequations}
\begin{align}
\gamma_1 x_1 + \gamma_2 x_2 = 1, \\
\gamma_2 x_{2N-1} + \gamma_1 x_{2N} = 0.
\end{align}
\end{subequations}
Inserting the expression for \(x_2\) in terms of \(x_{2N}\), for \(x_{2N-1}\) in terms of \(x_1\), and solving the resulting coupled set of linear equations, we obtain 
\begin{subequations}
\begin{align}
x_1 = \frac{1}{\gamma_1 + \gamma_2} \frac{(-\gamma_1 / \gamma_2)^{N} - 1}{(-\gamma_1/\gamma_2 )^{N} - (-\gamma_2/ \gamma_1)^{N}},
\\
x_{2N} = \frac{1}{\gamma_1 + \gamma_2} \frac{1 - (-\gamma_2 / \gamma_1)^{N}}{(-\gamma_1/\gamma_2)^{N} - (-\gamma_2/\gamma_1)^{N}}.
\end{align}
\end{subequations}
\noindent Using this together with Eq.~\eqref{eq_app_diffEqSol}, we obtain 
\begin{align}
\bm{v}^T A^{-1} B \bm{v} = 
\left( \frac{\gamma_1 - \gamma_2}{\gamma_1 + \gamma_2} \right) \frac{\gamma_1^{N} - (-\gamma_2)^{N}}{\gamma_1^{N} + (-\gamma_2)^{N}}.
\end{align}
\noindent The conductance is therefore
\begin{align}
G =  \frac{(g \mu_B)^2 }{\pi \hbar} \left(\frac{\sigma}{2n } \right) \frac{ (\sigma + 2n )^N  - (\sigma - 2n)^N}{(\sigma + 2n)^{N} + (\sigma - 2n)^{N} }.
\end{align}
\noindent As long as \(\sigma \neq 0\) and \(n\neq 0\), either \((\sigma + 2n )^N\) or \((\sigma -2n )^N\) can be neglected in the limit \(N \rightarrow \infty\). This gives the two-terminal spin conductance 
\begin{align}
G =\frac{(g \mu_B)^2 }{\pi \hbar} \left(\frac{\sigma}{2n } \right),
\end{align}
\noindent as also given in the main text.

\bibliographystyle{apsrev4-2}  
%\bibliography{ref}


\end{document}